\documentclass[showpacs,amsmath,amssymb,aps,twocolumn,longbibliography,pra]{revtex4-2}
\usepackage{graphicx}
\usepackage[colorlinks=true,citecolor=blue,linkcolor=magenta]{hyperref}
\usepackage[usenames]{color}
\usepackage{relsize}
\usepackage[english]{babel}
\usepackage{enumerate}
\usepackage{bbm}
\usepackage{xcolor}

\newcommand{\ket}[1]{\left| #1 \right>} 
\newcommand{\bra}[1]{\left< #1 \right|} 

\newcommand {\grsim} {\ {\raise-.5ex\hbox{$\buildrel>\over\sim$}}\ }
\newcommand {\lessim} {\ {\raise-.5ex\hbox{$\buildrel<\over\sim$}}\ } 
\newcommand {\ii} {i}

\usepackage{xcolor}
\usepackage{siunitx}

\newcommand{\RN}[1]{%
  \textup{\uppercase\expandafter{\romannumeral#1}}%
}

\usepackage{xr}
\newcommand{\nocontentsline}[3]{}
\newcommand{\tocless}[2]{\bgroup\let\addcontentsline=\nocontentsline#1{#2}\egroup}

\begin{document}

\title{Benchmarking multi-qubit gates - I: Metrological aspects}

\author{Bharath Hebbe Madhusudhana }
\author{}

\affiliation{$^{1}$\,Fakult\"at f\"ur Physik, Ludwig-Maximilians-Universit\"at M\"unchen, Schellingstra{\ss}e 4, 80799 M\"unchen, Germany}
\affiliation{$^{2}$\,Munich Center for Quantum Science and Technology (MCQST), Schellingstr. 4, 80799 M\"unchen, Germany}
\affiliation{$^{3}$\,Max-Planck-Institut f\"ur Quantenoptik, Hans-Kopfermann-Stra{\ss}e 1, 85748 Garching, Germany}


\begin{abstract} 
 Accurate and precise control of large quantum systems is paramount to achieve practical advantages on quantum devices. Therefore, benchmarking the hardware errors in quantum computers has drawn significant attention lately.  Existing benchmarks for digital quantum computers involve averaging the global fidelity over a large set of quantum circuits and are therefore unsuitable for specific many-qubit control operations used in analog quantum operations.  Moreover, average global fidelity is not the optimal figure-of-merit for some of the applications specific to analog devices, such as the study of many-body physics, which often use local observables. In this two-part paper,we develop a new figure-of-merit suitable for analog/multi-qubit quantum operations based on the reduced Choi matrix of the operation. In the first part, we develop an efficient, scalable protocol to completely characterize the reduced Choi matrix.  We identify two sources of sampling errors in measurements of the reduced Choi matrix and we show that there are fundamental limits to the rate of convergence of the sampling errors, analogous to the standard quantum limit and Heisenberg limit.  A slow convergence rate of sampling errors would mean that we need a large number of experimental shots.  We develop protocols using quantum information scrambling, which has been observed in disordered systems for e.g., to speed up the rate of convergence of the sampling error at state preparation Moreover, we develop protocols using squeezed and entangled initial states to enhance the convergence rate of the sampling error at measurement, which results in a metrologically enhanced reduced process tomography protocol.  
 
\end{abstract}

\maketitle

\tocless\section{Introduction}

\begin{figure}[h!]
\includegraphics[scale=0.34]{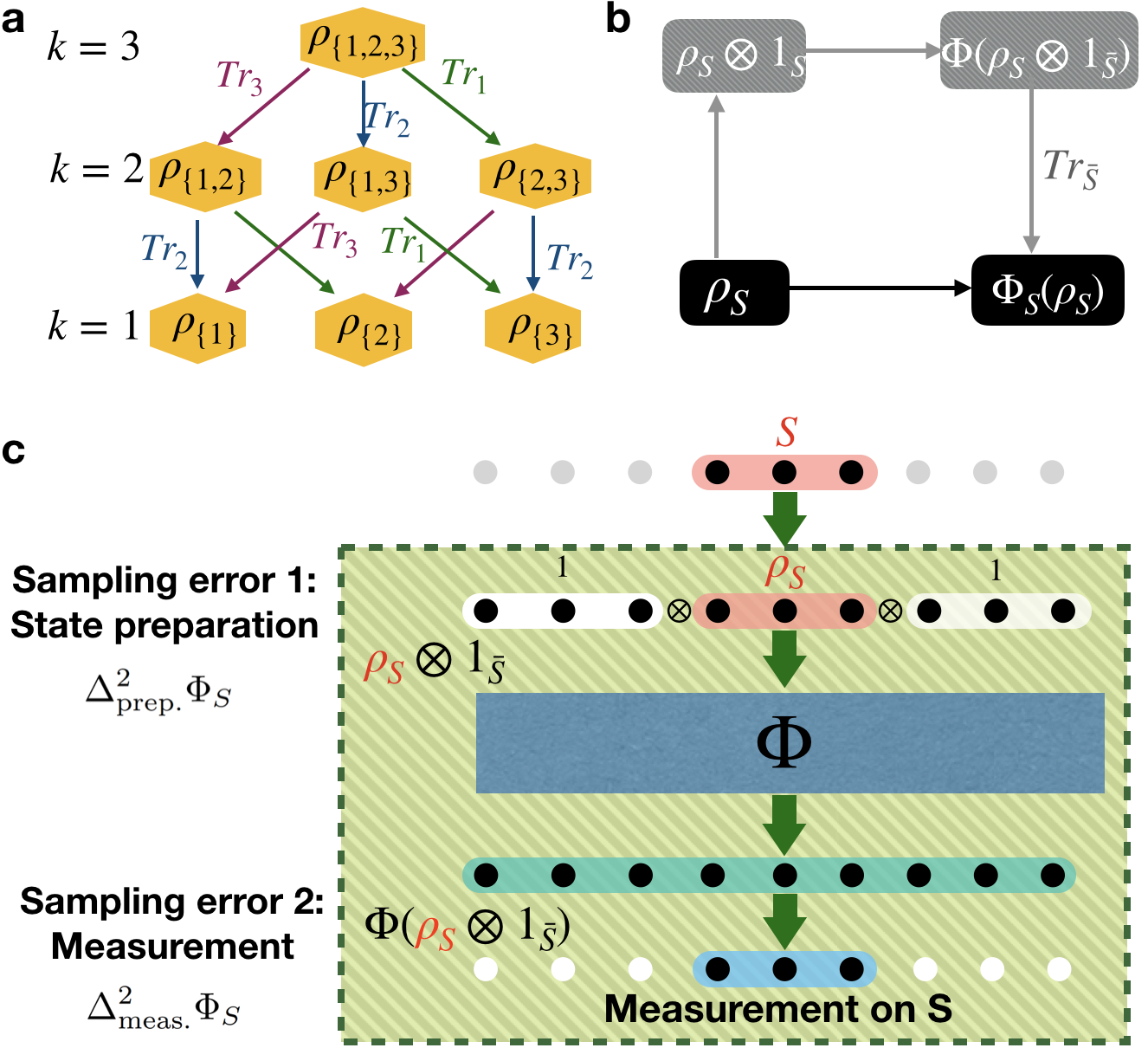}
\caption{\textbf{Reduced process: }\textbf{a. }The subsystems of a system of $N$ qubits can be arranged in a hierarchy of inclusion.   Corresponding to a state of the system,  for every subset $S\subset \{1, 2, 3, \cdots, N\}$, one can define a reduced density matrix.  These reduced density matrices form a Hierarchy under partial trace.  The various levels of the Hierarchy are characterized by number of qubits in the subsystem.  Figure shows a schematic of the hierarchy for $N=3$. \textbf{b.} Corresponding to a process $\Phi$ acting on the full system, one can define, for every subsystem $S$, a reduced process $\Phi_S$.  The reduced processes also form a Hierarchy.  \textbf{c.} shows a tomography of the reduced process. The measurement involves preparing a mixed initial state, resulting in two sources of sampling errors --- one at state preparation and one at measurement. }\label{FIG1}
\end{figure}
A principle challenge in developing quantum hardware is to reliably actualize a wide range of control operations on quantum systems, e.g., a set of qubits. This is an important challenge not only in the development of quantum computers and quantum simulators, but also in the development of other, non-computational quantum technologies such as quantum metrology and quantum communications.  Consequently, benchmarking quantum operations has been a vibrant area of research lately.

There are two fundamental challenges in benchmarking quantum operations. First, a quantum operation is characterized by an exponentially large number of variables~\cite{Chuang_1997} and therefore a process tomography, i.e., a complete characterization, is not scalable. And second, the expected effect of a quantum operation on the quantum system is sometimes outside the classical computational limits and \textit{desirably} so, making it impossible to have a reference to compare the experimental implementation of the operation with. The former is a quantum metrology problem --- a solution would involve designing a protocol which can be implemented without adding significant noise, in order to measure the desired parameters of the operation. The latter is a computational challenge --- a solution to it would involve finding efficiently verifiable properties of the operation.

\begin{figure*}[ht]
\includegraphics[scale=0.38]{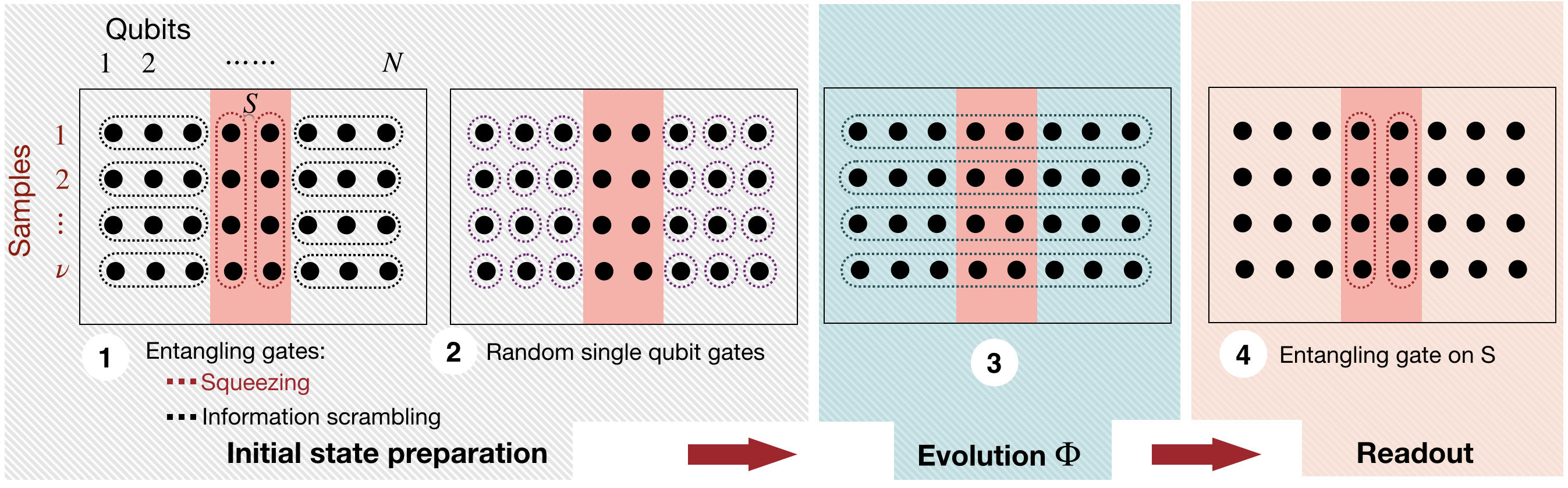}
\caption{\textbf{Metrologically enhanced reduced process tomography protocol}: We consider a $2$D array of $N\times \nu$ qubits ---  $\nu$ statistical repetitions of a system of $N$ qubits realized in parallel so that one can use metrological techniques to enhance the readout.  The protocol is designed for a reduced process tomography $\Phi_S$ of a subsystem $S$ of the $N$ qubits, corresponding to an $N$-qubit operation $\Phi$.  It consists of $4$ steps. In Step $1$, we apply entangling operations to $\bar{S}$ within each realization in order to scramble the information. This reduces the sampling error in state preparation $\Delta^2_{\text{prep}}\Phi_S$ (see Sec.~V).  This step also includes entangling gate on qubits in $S$ \textit{across} realizations. These are squeezing operations aimed at reducing the  sampling error in readout $\Delta^2_{\text{meas}}\Phi_S$ (see Sec.~VI). Step $2$ consists of independent random single qubit gates on $\bar{S}$ in order to prepare a uniform mixed state. Step $3$ consists of evolution under the target operation $\Phi$ and step $4$ involves a readout.  }\label{Fig2}
\end{figure*}


Some of the above challenges are addressed by randomized benchmarking protocol \cite{PRXQuantum.3.020357, Emerson_2005}, by considering cyclic quantum circuits. That is, quantum circuits that correspond to a unitary evolution of $\mathbbm{1}$. Ideally, such circuits map each state to itself. Therefore, one can start with an $N-$ qubit state $\ket{0}^N \in \mathcal H^{\otimes N}$, which can be prepared reliably, one can measure the fidelity of the final state with $\ket{0}^N$.  $\mathcal H$ is the Hilbert space of one qubit.  Ideally, one must average this fidelity over various initial states. However, arbitrary initial states cannot be prepared reliably on current devices. Assuming that the errors are independent of the gates applied, we can  average over cyclic circuits instead \cite{NIELSEN2002249}.  The simplest way to produce a cyclic circuit is to mirror a circuit --- applying a set of gates and inverting them. However, the typical depth of the circuit should be exponential in the system size, in order to effectively represent a Haar random unitary \cite{Pozniak_1998}. Moreover, this method overlooks some systematics, which maybe erased by the mirroring. A popular alternative is to use Clifford gates. That is, restricting the gates used in the circuits to Clifford gates \cite{PhysRevA.77.012307, PhysRevLett.106.180504}. Circuits composed of Clifford gates are classically simulable. Moreover, the number of gates necessary to produce a  typical Clifford gate is only polynomial. However, it is unclear whether a Clifford gate benchmark provides a useful estimate of the average fidelity, given that a typical unitary gate consists of many more gates than a typical Clifford operation. This idea has been extended to other groups, besides the Clifford group as well~\cite{cs}. Recently, there have been works combining the idea of circuit mirroring and Clifford gates \cite{Proctor_2021} to tailor the benchmarking scheme to target errors of a specific nature.  Alternatively, one can consider averaging over a subgroup of $SU(2^N)$~\cite{Emerson_2007}. Recently, a new protocol to benchmark a \textit{specific} unitary $U$ as opposed to averaging over several circuits, was proposed based on the symmetries of $U$~\cite{PhysRevLett.123.060501}.  Another approach to benchmarking a specific unitary is to use random matrix theory and test the expected statistical properties such as moments of the output distribution.  For instance,  a new benchmarking method based on emergent Porter-Thomas distributions in the output state after time evolution under a specific many-body Hamiltonian was recently developed~\cite{arxiv.2103.03536, arxiv.2103.03535}. While this method is suitable for specific many-qubit operations, it does not scale efficiently with the system size.    


Here,  we develop a new figure-of-merit and a protocol to benchmark the experimental implementation of a given unitary $U\in SU(2^N)$, which is produced either by a circuit $C$ consisting of one and two qubit gates or by time evolution under a many-qubit Hamiltonian $H$. The former is relevant for digital quantum computers built using ion traps/superconducting circuits/neutral atoms and $U$ would be the ordered product of the unitaries corresponding to the gates in $C$. The latter is relevant for analog quantum simulators built for e.g., using trapped neutral atoms and $U=e^{-\ii Ht}$ where $t$ is the duration of the time evolution.  We focus on this case in this work. 

This is a desirable goal for applications  such as quantum certified approximations, where one uses an analog quantum computer to benchmark the performance of a new classical approximation ansatz,  shown in the recent work~\cite{PRXQuantum.2.040325}.  In fact, one can advance this idea further --- train a classical neural network on the data from a quantum simulator in order to develop implicit classical approximations~\cite{Huang_2022, Huang_2022_2}.  These applications only need a few accurate many-qubit operations. Moreover,  quantum computers based alternate gate-sets that include direct application of many-qubit operations have been studied recently~\cite{https://doi.org/10.48550/arxiv.2210.02936, Qsim_review}.  One such example is a digital-analog quantum computer, where universal control is achieved using a combination of a few many-qubit operations and single-qubit gates~\cite{PhysRevA.101.022305, PhysRevResearch.2.013012}. Experimental platforms are also being developed~\cite{Yu_2022} where, our present goal of benchmarking a specific many-qubit operation would be very relevant.\\

\tocless\section{Results}

Every benchmarking protocol is anchored to a figure-of-merit --- the quantity which characterizes the quality of the quantum operation and which we intend to measure through the protocol. The figure-of-merit is chosen carefully, tailored to a desired application of the quantum device.  The existing benchmarking protocols use the global fidelity,  averaged over a large set of circuits as the figure-of-merit and therefore evaluate the entire device as a whole, as opposed to what we need --- a figure-of-merit that evaluates the accuracy of implementation of the \textit{specific} unitary $U$.  Moreover, the the global fidelity is not always the relevant measure.  An $N-$qubit state contains a large volume of quantum information, with an intricate structure. One can organise this information in a hierarchy, where the lowest strata consists of reduced density matrices for each qubit and the higher strata consist of correlations of various orders among subsets of the $N$ qubits (Fig.~\ref{FIG1}a).  While the global fidelity between two quantum states represents an aggregation of the errors incurred at various strata of the hierarchy of quantum information, it is \textit{one} number, making it hard to extract the component of the error we may be specifically targeting.  One can define fidelities of reduced density matrices of various subsets. These fidelities represent errors coming form various sources.  Building up an analogy between states and quantum processes (see. Fig.~\ref{FIG1}), we define a \textit{reduced process tomography} i.e., tomography of a process restricted to a small subset of the qubits in order to develop a benchmarking protocol that addresses errors specific to the chosen subset Fig.~\ref{FIG1}b. If $S\subset \{1, 2, \cdots, N\}$ is a subset of the system and $\Phi$ is a quantum process acting on the whole system, we can define reduced process on the subset $S$ as $\Phi_S (\rho_S) = \text{Tr}_{\bar{S}}(\Phi(\rho_S \otimes \frac{1}{2^{|\bar{S}|}}\mathbbm{1}_{\bar{S}}))$. Here, $\bar{S}=\{1, 2, \cdots, N\}-S$ and $|\bar{S}|$ is the number of qubits in it.

The initial state in a reduced process tomography is necessarily mixed and is produced using controlled samples that average to the target mixed state. Therefore, we have two independent sources of sampling errors in such a tomography measurement --- one corresponding to the initial state and the other corresponding to the measurement (Fig.~\ref{FIG1}c). Therefore, the total sampling error is given by
\begin{equation}\label{total_error}
\Delta^2 \Phi_S = \Delta^2_{\text{prep.}} \Phi_S +  \Delta^2_{\text{meas.}} \Phi_S 
\end{equation}
See ref.~\cite{suppmat} for details and explanation for this expression. Most sampling errors with $\nu$ uncorrelated samples scale as $ \Delta^2_{\text{prep.}} \Phi_S \sim \frac{1}{\nu}$ (and $\Delta^2_{\text{meas.}} \Phi_S \sim \frac{1}{\nu}$).  Thus, the total square error also scales inversely with $\nu$.  There are two main problems pertaining to benchmarking via reduced process tomography:
\begin{itemize}
\item[1.] \textbf{Metrological aspects:} Enhancing the convergence rate of the sampling errors in the tomography measurements.
\item[2.] \textbf{Computational aspects:} Developing benchmarks using the reduced process $\Phi_{S}$.
\end{itemize}
In this paper, we address the first of the above two problems. We develop protocols to enhance the convergence of  the total sampling error. In section V and VI, we develop entanglement-based protocols to speed up the convergence of the sampling error in state preparation, i.e., in $\Delta^2_{\text{prep.}} \Phi_S $ and in measurement, i.e., in $\Delta^2_{\text{meas.}} \Phi_S $. In section III and IV, we develop from background material.  The computational aspect will be addressed in the part II of this paper~\cite{BHM2}.  \\

\tocless\section{The reduced Choi matrix}
We refer to the experimental implementation of $U$, by the map $\Phi$. In order to maintain generality, we model this operation by a the most general quantum operation, i.e., a completely positive map. The Choi matrix corresponding to this map is a $4^N \times 4^N$ matrix $\rho^{\Phi}$, defined on the space $\mathcal H^{\otimes N}\otimes \mathcal H^{\otimes N}$ as $\rho^{\Phi}_{ij; kl} = \text{Tr}(\Phi(\ket{i}\bra{j})\ket{k}\bra{l})$. Here, $\ket{i}, \ket{j}, \ket{k}$ and $\ket{l}$ are basis elements of $\mathcal H^{\otimes N}$.  One can view $\rho^{\Phi}$ as a block matrix, where the $i,j-$th block is the $2^N\times 2^N$ matrix $\Phi(\ket{i}\bra{j})$.  Moreover, If we choose $\rho$ as an initial state and measure $\hat{O}$ after the quantum operation $\Phi$, the expectation value is given by $\text{Tr}(\hat{O}\Phi(\rho)) = \text{Tr}(\rho^{\Phi}\hat{O}\otimes \rho)$ (see ref.~\cite{suppmat} for more details). The unitary $U$ has its own Choi matrix representation $\rho^U$.  For a given subset of qubits $S\subset \{1, 2, \cdots, N\}$, we define the reduced Choi matrix as the partial trace
\begin{equation}
\rho^{\Phi, S} = \text{Tr}_{\bar{S}} \rho^{\Phi}
\end{equation}
Here, $\bar{S}=\{1, 2, \cdots, N\}-S$.  Similarly, the reduced Choi matrix for the unitary $U$ is $\rho^{U, S} = \text{Tr}_{\bar{S}}\rho^{U}$.  If $S$ contains $m$ qubits, the reduced Choi matrix is a $4^m\times 4^m$ matrix that represents the effect of the time evolution on the subset $S$. To obtain a precise interpretation, if $\hat{O}$ is an observable acting on $S$ and $\rho_S$ is a state of $S$,  it follows that 
\begin{equation}\label{reduced_choi}
\text{Tr}(\rho^{\Phi, S}\hat{O}\otimes \rho_S) = \text{Tr}\left[\hat{O}\otimes \mathbbm{1}_{\bar{S}}\Phi\left(\rho_S\otimes\frac{1}{2^{N-m}}\mathbbm{1}_{\bar{S}}\right)\right]
\end{equation}
That is, the partial trace represents the process on $S$, that $\Phi$ would induce on a given state $\rho_S$ of $S$, with the rest of the qubits \textit{initially} in the uniformly mixed state $\frac{1}{2^{N-m}}\mathbbm{1}_{N-m}$.  Note that if $U$ is an entangling operation, the partial trace $\rho^{U, S}$ can be quite general and non-trivial.  The Choi matrix of a quantum operations has all the properties of a quantum state in a higher dimensional Hilbert space and therefore, a partial trace is a natural choice for reduction (See ref.~\cite{suppmat} for more details). 

The reduced Choi matrix would be affected my most errors that would also affect the reduced density matrix of $S$, after the time evolution.  Moreover, for small $m$ (we use $m=1, 2$ below), it is possible to experimentally measure the reduced Choi matrix. Therefore, it makes a good candidate that can be used to benchmark the time evolution. Eq.~\ref{reduced_choi} provides a way of measuring the reduced Choi matrix.  One can obtain   $\text{Tr}(\rho^{\Phi, S}\hat{O}\otimes \rho) $ by preparing $S$ in the state $\rho_S$ and the rest of the qubits in the uniform mixed state and measuring $\hat{O}$ on $S$, after the quantum operation.  By varying $\hat{O}$ and $\rho$, one can reconstruct the reduced Choi matrix $\rho^{\Phi, S}$.  We refer to this process as the r\textit{educed process tomography}. We can then compare $\rho^{\Phi, S}$ with $\rho^{U, S}$ to benchmark the operation (the details on the efficiently verifiable properties of $\rho^{U, S}$ will be presented in part-II of this paper ). \\

\tocless\subsection{Reduced process tomography}
The tomography of $\rho^{\Phi, S}$ is straightforward. For instance, consider $m=1$, i.e., $S=\{1\}$. $\rho^{\Phi, S}$ is a $4\times 4$ matrix and has $16$ free parameters. One can choose 
\begin{equation}
\begin{split}
\hat{O}, \rho_S \in \mathcal T =& \left\{ \ket{0}\bra{0},  \ket{1}\bra{1}, \frac{1}{2}(\ket{0}+\ket{1})(\bra{0}+\bra{1}), \right.\\
& \left. \frac{1}{2}(\ket{0}+i\ket{1})(\bra{0}-i\bra{1})\right\} 
\end{split}
\end{equation}. There are $16$ such combinations of $\rho_S, \hat{O}$. One can reconstruct $\rho^{\Phi, S}$ using these $16$ measurements. This idea extends to $m>1$, with $\rho_S, \hat{O}\in \mathcal T^m$. However, it is practically challenging to go beyond $m=2$, although the protocol we present below is scalable in $N$.  

The natural protocol, suggested by Eq.~\ref{reduced_choi}.  For  each pair $\tau_1, \tau_2 \in \mathcal T^m$,  the component of $\rho^{\Phi, S}$  is
\begin{equation}\label{exp_tau1_tau2}
\rho^{\Phi, S}_{\tau_1,\tau_2} =  \text{Tr}\left[\tau_2\otimes \mathbbm{1}_{\bar{S}}\Phi\left(\tau_1\otimes\frac{1}{2^{N-m}}\mathbbm{1}_{\bar{S}}\right)\right]
\end{equation}
Therefore, we prepare the $m$ qubits in $S$ in the state $\tau_1$, the remaining $N-m$ qubits in the  uniform mixed state, $\frac{1}{2^{N-m}}\mathbbm{1}_{N-m}$ and apply the operation $\Phi$, followed by a measurement of $\tau_2$ on the $m$ qubits in $S$. The expectation value of this measurement is the component $\rho^{\Phi, S}_{\tau_1,\tau_2} $. Two questions remain: what is the convergence rate of the measurement to the expectation value, i.e., how many experimental shots do we need? and how do we prepare the $N-m$ qubits in the uniform mixed state, $\frac{1}{2^{N-m}}\mathbbm{1}_{\bar{S}}$? The two are connected --- the convergence rate also depends on the error induced in the preparation of the uniformly mixed state. In other words, one can optimize the preparation strategy to maximize the convergence rate. 

Although, it is apparent that preparing a quantum system in the uniform mixed state is straightforward, it is experimentally quite challenging to ensure scalability with $N$.  An incoherent sum of Haar random states quickly converges to the uniform mixed state for arbitrary $N$. However, a typical Haar random state is highly entangled and in a real experiment, we can only reliably prepare states with very low entanglement.

We consider a simple example to illustrate the problem. We can produce a single qubit uniformly mixed state $\frac{1}{2}\mathbbm{1}$ using a set of $\nu$ experimental shots, where in each shot the qubit is prepared randomly in $\ket{0}$ or $\ket{1}$ with equal probability. We refer to the state prepared after $\nu$ shots as $\rho_{\text{real}, \{1\}} = 1/\nu\sum_{i=1}^{\nu}\ket{z_i}\bra{z_i} $ (the $\{1\}$ indicates we are in a single qubit system).  Here, $z_i=0, 1$.  The Uhlmann fidelity between $\rho_{\text{real}, \{1\}}$ and $\frac{1}{2}\mathbbm{1}$ is
\begin{equation}
F\left(\rho_{\text{real}, \{1\}}, \frac{1}{2}\mathbbm{1}\right)=\frac{1+2\sqrt{s(1-s)}}{2}
\end{equation}
Here, $s=\frac{1}{\nu}\sum_i z_i$.  $s$ has an average value of $0$ and a standard deviation of $\frac{1}{2\sqrt{\nu}}$.  Thus,  the fidelity has an average of $1-1/4\nu$. If we now extend this to $N$ qubits, i.e., pick the state of each qubit to be $\ket{0}$ or $\ket{1}$ at random, for each shot $\nu$, the prepared state is $\rho_{\text{real}} = \frac{1}{\nu}\sum_{i=1}^{\nu}\bigotimes_{j=1}^N \ket{z_{ij}}\bra{z_{ij}}$, where each $z_{ij}=0, 1$ chosen at random.  It is straihghtforward to show that the Uhlmann fidelity is 
\begin{equation}
F\left(\rho_{\text{real}}, \frac{1}{2^N}\mathbbm{1}\right)=\Pi_{j=1}^N \frac{1+2\sqrt{s_j(1-s_j)}}{2}
\end{equation}
Here, $s_j =\frac{1}{\nu}\sum_i z_{ij}$.  If the state of the qubit are assumed to independent random variables,  the average fidelity is $(1-1/4\nu)^N$.  Note the exponential decay. Below, we will formalize this result into a theorem and also show a quantitative relation between the entanglement of the states and the convergence rate.  Before that, we will address the important question as suggested by the intriguing exponential scaling of the Uhlmann fidelity: \textit{what is the most appropriate measure of distance to use in the convergence analysis}? \\

\tocless\subsection{Characterising the error in mixed state preparations}
The most logical way of deciding on a distance measure to evaluate the error is using the intended \textit{purpose} for which the mixed state. is being prepared. In this case, the purpose is to measure $\rho^{\Phi, S}_{\tau_1,\tau_2}$ and therefore, we will pick a measure of the distance that is induced by the maximum error in $\rho^{\Phi, S}_{\tau_1,\tau_2}$.  Looking at Eq.~\ref{reduced_choi}, the quantity of interest is linear in the state of the $N-m$ qubits in $\bar{S}$. Indeed, if one experimentally prepares $\rho_{\text{real}, \bar{S}}$ while attempting to prepare $\frac{1}{2^{N-m}}\mathbbm{1}_{\bar{S}}$, we would obtain
\begin{equation}
\tilde{\rho}^{\Phi, S}_{\tau_1,\tau_2} =  \text{Tr}\left[\tau_2\otimes \mathbbm{1}_{N-m}\Phi\left(\tau_1\otimes\rho_{\text{real}, \bar{S}}\right)\right]
\end{equation}
and the error would be
\begin{equation}
\Delta \rho^{\Phi, S}_{\tau_1, \tau_2}= |\rho^{\Phi, S}_{\tau_1,\tau_2} - \tilde{\rho}^{\Phi, S}_{\tau_1,\tau_2}| =|  \text{Tr}\left[\tau_2\otimes \mathbbm{1}_{\bar{S}}\Phi\left(\tau_1\otimes\epsilon \right)\right]|
\end{equation}
where $\epsilon = \frac{1}{2^{N-m}}\mathbbm{1}_{\bar{S}}-\rho_{\text{real}, \bar{S}}$. We may rewrite this as 
\begin{equation*}
\Delta \rho^{\Phi, S}_{\tau_1, \tau_2} = |\text{Tr}(\rho^{\Phi} \tau_2\otimes \mathbbm{1}_{\bar{S}} 
\otimes \tau_1\otimes \epsilon)| = |\text{Tr}(P^{\Phi}_{\tau_1, \tau_2}\epsilon)|
\end{equation*}
Here, $P^{\Phi}_{\tau_1, \tau_2}$ a $2^{N-m}\times 2^{N-m}$ matrix, given by the partial trace of $\rho^{\Phi} \tau_2\otimes \mathbbm{1}_{N-m} 
\otimes \tau_1\otimes \mathbbm{1}_{\bar{S}}$, with respect to all qubits excluding the ones in $\bar{S}$.  Cauchy-Schwarz inequality reads
\begin{equation}
\Delta \rho^{\Phi, S}_{\tau_1, \tau_2}  = |\text{Tr}(P^{\Phi}_{\tau_1, \tau_2}\epsilon)| \leq ||\epsilon||_2 ||P^{\Phi}_{\tau_1, \tau_2}||_2
\end{equation}
Here, $||\cdot ||_2$ is the Frobenius norm or the Hilbert-Schmidt norm or, equivalently, the Schatten-2 norm defined as $||X||_2 =\sqrt{\text{Tr}(X^{\dagger}X)} = \sqrt{\sum_{i,j}|X_{ij}|^2}$.  Thus,  the norm $||\epsilon||_2$ is a suitable measure of error.  In case of fixed $\epsilon$, i.e., a systematic error, this quantity is straightforward to interpret as the pythogorean length of the error.  However,  in real experiments,  $\epsilon$ is a random matrix and therefore, we need to construct a classical average of the error measure. The obvious candidate to quantify the error is $\langle ||\epsilon||_2^2\rangle $, where $\langle \cdot \rangle$ represents averaging over the classical statistical distribution of $\epsilon$, to account for the fact that mixed states are produced by a statistical average of pure states.  This was done in the example above, when we obtained an average fidelity of $1-1/\nu$. We will however show that this is not the most appropriate way of averaging the error measure. 

To identify the most logical way of averaging the error, we gain turn to to the \textit{purpose} of estimating this error --- we need to know the average of $\text{Tr}(P^{\Phi}_{\tau_1, \tau_2}\epsilon)$.  We invoke the covariance matrix of $\epsilon$ here. The fluctuation of the random matrix $\epsilon$ is characterized by the covariance matrix $\chi$ of $\epsilon$, which is the $4^{N-m}\times 4^{N-m}$ matrix defined as
\begin{equation*}
\chi_{ij; kl} = \langle \epsilon_{ij}\epsilon_{lk} \rangle
\end{equation*}
Here, $\langle \cdot \rangle$ represents the classical average over the random $\epsilon$.   It follows that $\langle \epsilon^2\rangle$ is a partial trace of $\chi$. 
\begin{equation*}
\langle \epsilon^2\rangle _{ij} = \sum_k \chi_{ik;jk}
\end{equation*}
and
\begin{equation}\label{eps_to_chi}
\langle ||\epsilon||_2^2\rangle = \langle \text{Tr}(\epsilon^2)\rangle = \text{Tr}(\chi)
\end{equation}
The quantity of interest is the average error in measurements, given by the square average of  $\text{Tr}( P^{\Phi}_{\tau_1, \tau_2} \epsilon)$
\begin{equation*}
\begin{split}
\langle \Delta^2 \rho^{\Phi, S}_{\tau_1, \tau_2}\rangle= &\langle |\text{Tr}(P^{\Phi}_{\tau_1, \tau_2} \epsilon )|^2\rangle \\
=&\sum \chi_{ij;kl}[P^{\Phi}_{\tau_1, \tau_2}]_{ij}[P^{\Phi}_{\tau_1, \tau_2}]_{kl} \leq \sigma_{\text{max}}(\chi)||P^{\Phi}_{\tau_1, \tau_2}||_2^2
\end{split}
\end{equation*}
Thus, the most relevant measure of error is given by the maximum singular value of $\chi$, $\sigma_{\text{max}}(\chi)$.  The average of the Frobenious norm, $\langle ||\epsilon||_2^2\rangle $, is in fact the sum of the singular values of $\chi$ (Eq.~\ref{eps_to_chi}) and is therefore much larger than the above limit. 
\begin{equation*}
\sigma_{\text{max}}(\chi) \leq \langle \text{Tr}(\epsilon^2)\rangle = \text{Tr}(\chi)
\end{equation*}
The equality holds for \textit{fixed} $\epsilon$.  Therefore, we use $\sigma_{\text{max}}(\chi)$ as a measure of the convergence rate.  See ref.~\cite{suppmat} for a detailed discussion of the meaning of Uhlmann fidelity and why it is not the most suitable measure for our purposes. In the next section, we present a theorem to on the convergence rate of the error $\langle \Delta^2 \rho^{\Phi, S}_{\tau_1, \tau_2} \rangle $ to zero and how it depends on the entanglement of the initial states, analogous to the standard quantum limit and Heisenberg limit in quantum metrology.\\

\tocless\section{Convergence rate and its fundamental limits}\label{limits_on_rate}
There are two extreme ways to prepare the qubits in $\bar{S}$ in the uniform mixed state. One is to prepare the qubits in $\bar{S}$, in each experimental shot, in a Haar-random state in $\mathcal H^{\otimes (N-m)}$. The other is to prepare each individual qubit in $\bar{S}$ in an independent Haar random state. The latter is always a product state, while the former is most likely to be a highly entangled state.  We consider an intermediate method: divide the $N-m$ qubits in $\bar{S}$ into groups of $\ell$ qubits each and prepare each group in an independent Haar random state from $\mathcal H^{\otimes \ell}$. The qubits within each group will most likely be highly entangled, while the states of two different groups will always be separable. 
We refer to such a state as a \textit{Haar random $\ell-$bit state}. When $\ell =N-m$, this reduces to one of the extreme methods described above and when $\ell =1$, it reduces to the other.  The group size $\ell$ is a measure of entanglement in the state and we will show that the convergence depends on this.  We state the main theorem of this section now

\textbf{Convergence theorem - 1: } Let us assume we use $\nu$ experimental shots with an initial state $\tau_1$ for the qubits in $S$ and Haar random $\ell-$bit states for qubits in $\bar{S}$ and measure $\tau_2$ after the operation $\Phi$. The state of $\bar{S}$ converges to $\frac{1}{2^{N-m}}\mathbbm{1}_{\bar{S}}$ at the rate
$$
\sigma_{\max}(\chi(\bar{S})) = \frac{1}{\nu 2^{N-m}(2^{\ell}+1)}
$$
An, we can estimate $\rho^{\Phi, S}_{\tau_1, \tau_2}$ with a precision of 
\begin{equation}\label{thm_1}
\sqrt{\langle \Delta^2 \rho^{\Phi, S}_{\tau_1, \tau_2} }\rangle \leq \frac{||P^{\Phi}_{\tau_1, \tau_2}||_2}{\sqrt{\nu} \sqrt{2^{N-m}(2^{\ell}+1)}} \approx \frac{1}{\sqrt{\nu} \sqrt{(2^{\ell}+1)}}
\end{equation}

\noindent\textbf{Proof:} We provide a sketch of the proof of this theorem for $N-m$ qubits and a rigorous proof for $N$ \textit{qudits} in the supplementary information.  Let us begin with some observations for $N-m=1$ qubit.  We may write a density matrix $\rho$ in the basis $\left\{\frac{\mathbbm{1}}{\sqrt{2}}, \frac{\sigma_x}{\sqrt{2}}, \frac{\sigma_y}{\sqrt{2}}, \frac{\sigma_z}{\sqrt{2}}\right\}$. We have used the factor $1/\sqrt{2}$ to make each of the normalized in the sense of a vector.  For convenience, we use $\sigma_0=\frac{\mathbbm{1}}{\sqrt{2}}$, $\sigma_1=\frac{\sigma_x}{\sqrt{2}}$, $\sigma_2=\frac{\sigma_y}{\sqrt{2}}$ and $\sigma_3=\frac{\sigma_z}{\sqrt{2}}$.  A density matrix can be written as $\rho = v_0 \sigma_0 +\cdots +v_3\sigma_3$.  It follows from $\text{Tr}(\rho)=1$ and $\text{Tr}(\rho^2)=1$ for pure states, that $v_0=1/\sqrt{2}$ and $\sum_i v_i^2 =1$.  Also, $\rho-\frac{\mathbbm{1}}{2}=v_1\sigma_1 +v_2\sigma_2 +v_3\sigma_3$.  The covariance matrix $\chi$ averaging over random pure states $\rho$ is
$\chi = \text{diag}(0, \langle v_1^2\rangle , \langle v_2^2\rangle , \langle v_3^2\rangle )$.  Using the symmetry of the distribution and using $v_0^2 +\cdots +v_3^2=1$, it follows that $\langle v_i^2\rangle =1/6$ for $i=1, 2,3$.  Thus, the maximum singular value is $\sigma_{\text{max}}(\chi)=1/6$ and after $\nu$ uncorrelated repetitions, $\sigma_{\text{max}}(\chi)=\frac{1}{6\nu}$.

Let us now consider $\ell$ qubits. We can now construct a basis $\{\sigma_{\mathbf{i}}\}$ where $\mathbf{i}\in\{0, 1, 2, 3\}^\ell$ is a string of length $\ell$ with characters from $\{0, 1, 2, 3\}$. For instance, $\sigma_{1000}=\sigma_1\otimes\sigma_0^{\otimes 3}$.  Any density matrix can be written as $\rho = \sum_{\mathbf{i}} v_{\mathbf{i}}\sigma_{\mathbf{i}}$. Again, it follows that $v_{0^{\ell}} = \frac{1}{\sqrt{2^{\ell}}}$  and $\sum_{\mathbf{i}}v_{\mathbf{i}}^2 = 1$.  Thus, the covariance matrix $\chi$ of $\rho -\frac{\mathbbm{1}}{2^{\ell}}$ is given by $\chi = \text{diag}(0, \langle v_{\mathbf{i}}^2\rangle, \cdots)$. Again, using the symmetry, it follows that $\langle v_{\mathbf{i}}^2\rangle = \frac{1}{2^{\ell}(2^{\ell}+1)}$. Thus,  one of the singular values of covariance matrix is zero and the rest are $\frac{1}{2^{\ell}(2^{\ell}+1)}$.  After $\nu$ repetitions, $\sigma_{\text{max}}(\chi)=\frac{1}{\nu 2^{\ell}(2^{\ell}+1)}$. Note that the expression in the above theorem reduced to this when $N-m=\ell$. 

We will now consider the general case.  Let us assume that we have $r$ groups of $\ell$ qubits with $r=(N-m)/\ell$.  An $\ell$-bit state can be written as $\rho_1\otimes \cdots \otimes \rho_r$.  The covariance matrix of the tensor product of vectors, under independent distributions is indeed the tensor product of the covariance matrices of the individual vectors. However,  in order to apply this rule, we need the covariance matrix of $\rho_i$ and not $\rho_i -\frac{\mathbbm{1}}{2^{\ell}}$.  It is straightforward to see  that the covariance matrix of $\rho_i$ under Haar random states is $M=\text{diag}(\frac{1}{2^{\ell}}, \langle v_{\mathbf{i}}^2\rangle, \cdots) = \text{diag}(\frac{1}{2^{\ell}}, \frac{1}{2^{\ell}(2^{\ell}+1)} \cdots,  \frac{1}{2^{\ell}(2^{\ell}+1)})$. Note the strong similarity between $\chi$ and $M$. One of its singular values is $1/2^{\ell}$ and the rest are $\frac{1}{2^{\ell}(2^{\ell}+1)} $. The covariance matrix of $\rho_1\otimes \cdots \otimes  \rho_r$ would then be $M\otimes M\otimes \cdots \otimes M$.  We need the covariance matrix $\chi$ of $\rho_1\otimes \cdots \otimes \rho_r - \frac{\mathbbm{1}}{2^{N-m}}$, which is indeed very similar to $M^{\otimes r}$, except that the first diagonal term is zero in $\chi$ instead of $1/2^{N-m}$.  Thus,  $\sigma_{\max}(\chi) = \frac{1}{2^{\ell \times (r-1)}}\times \frac{1}{2^{\ell}(2^{\ell}+1)} = \frac{1}{2^{N-m}(2^{\ell}+1)}$. Convergence theorem-1 now follows $\blacksquare$

Note the extremes in Eq.~\ref{thm_1}: when $\ell=1$, the error is $1/\sqrt{6\nu}$ and when $\ell=N-m$,  it is $1/\sqrt{\nu(2^{N-m}+1)}$.  The parameter $\ell$ can also be considered as the entanglement depth of the initial states and this theorem is an analogue of the results presented in~\cite{Vitagliano_2017, PhysRevLett.107.180502}.  However,  one can not yet attribute the faster convergence to higher entanglement because, this theorem excludes classically correlated mixed states. That is, we have assumed no classical correlation between the $\ell-$bit states. For instance, one could use a correlated probability distribution to pick the $\ell-$bit states rather than using an uncorrelated Haar distribution. This raises a pertinent question: Is it possible to improve the convergence rate of the error if we classically correlate the $\ell-$bit states in different groups? The below theorem provides a negative answer to this question, thus establishing entanglement as an indispensable resource to improve convergence.  

\begin{figure}[h!]
\includegraphics[scale=0.37]{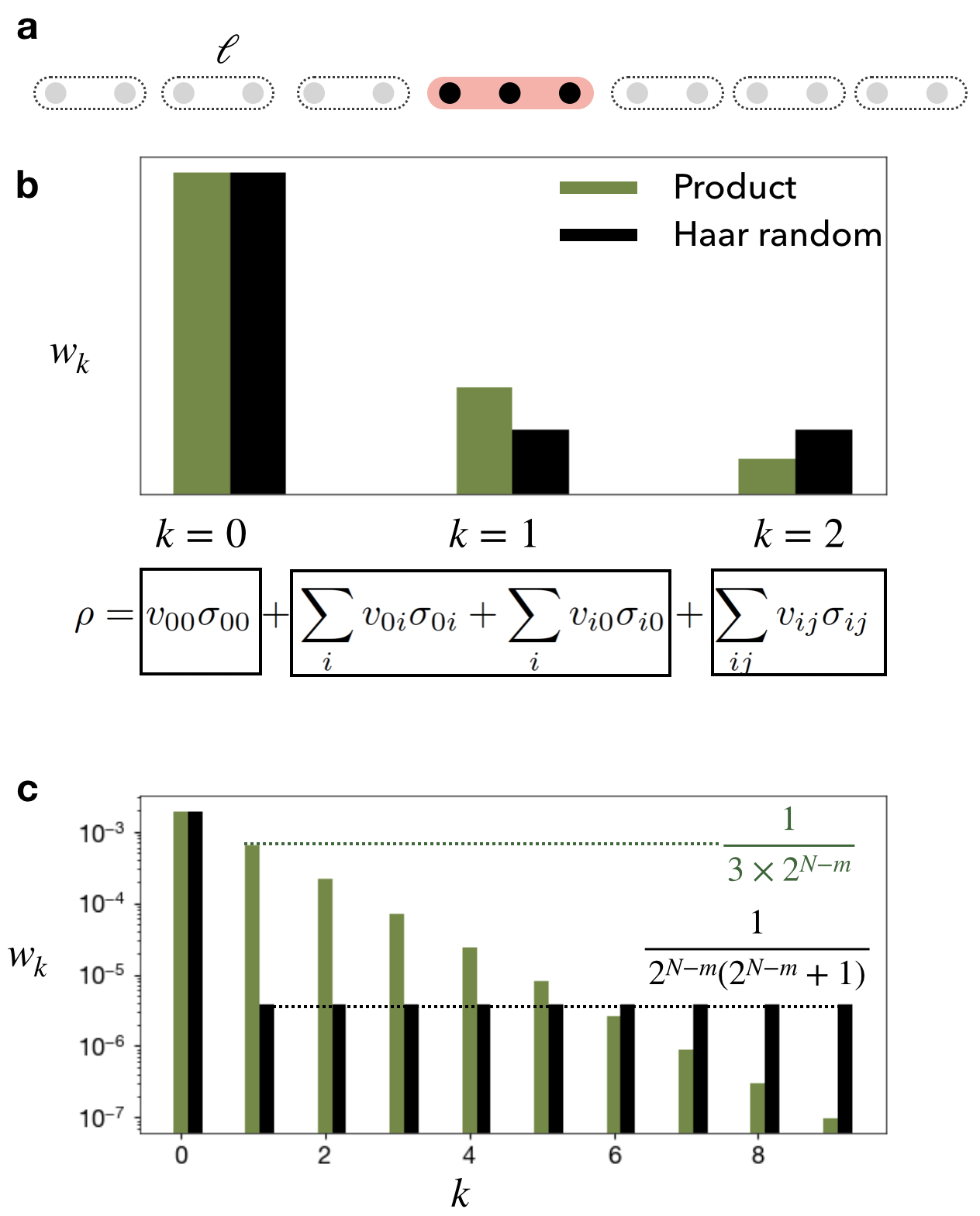}
\caption{\textbf{Distribution of information across the Hierarchy:} \textbf{a. } illustrates $\ell-$bit state of $\bar{S}$ obtained by dividing the $N-m$ qubits in $\bar{S}$ into groups containing $\ell-$bits each. An $\ell-$bit state is entangled within eaxch group, but separable between groups. \textbf{b. }shows the hierarchy of basis elements $\sigma_{\mathbf{i}}$ (see text) of density matrices.  The $k-$th level of the hierarchy consists of Pauli products $\sigma_{\mathbf{i}}$ that act non-trivially on $k$ qubits.  A Haar-random state has equal weight $w_k$(see text) at each level of the Hierarchy starting from $k=1$, whereas a product state has higher weights on lower levels of the Hierarchy.  \textbf{c.} shows the weight distribution across the hierarchy for Haar-random and separable states with $N-m=9$ qubits.}\label{FIG3}
\end{figure}

\textbf{Convergence theorem - 2:} If $\rho_{\text{real}}$ is a state of $N-m$ qubits produced using $\nu$ classically correlated samples of $\ell-$bit states with the goal of producing the uniformly mixed state, then the corresponding error $\sigma_{\text{max}}(\chi(\bar{S}))$ satisfies
\begin{equation}
\sigma_{\max}(\chi(\bar{S})) \geq \frac{1}{\nu 2^{N-m}(2^{\ell}+1)}
\end{equation}
Here, as usual, $\chi_{ij; kl}=\langle \epsilon_{ij}\epsilon_{kl}\rangle $ and $\epsilon=\frac{1}{2^{N-m}}\mathbbm{1}_{\bar{S}} -\rho_{\text{real}}$. 

\noindent\textbf{Proof:} We go back to the notation used in the proof of convergence theorem $1$.  If $\rho_1\otimes \cdots \otimes \rho_r$ is an random $\ell$ bit state, we need to compute the covariance matrix of $\rho_1\otimes \cdots \otimes \rho_r - \frac{\mathbbm{1}}{2^{N-m}}$, under a potentially correlated distribution over $\rho_i$. We cannot compute the tensor product of the individual covariance matrices here and therefore, we shall use a different approach. Let $\rho_i = \sum_{\mathbf{j}} v_{\mathbf{j}}^{(i)}\sigma_{\mathbf{j}} $. Let us consider the coefficient of $\sigma_{\mathbf{j}}\otimes \sigma_{0^{\ell}}\otimes \sigma_{0^{\ell}}\otimes \cdots \otimes \sigma_{0^{\ell}}$ in $\rho_1\otimes \cdots \otimes \rho_r - \frac{\mathbbm{1}}{2^{N-m}}$. It is $v_{\mathbf{j}}^{(1)}v_{0^{\ell}}^{(2)}\cdots v_{0^{\ell}}^{(r)} = v_{\mathbf{j}}^{(1)} \frac{1}{\sqrt{2^{\ell(r-1)}}}$.  Regardless of the underlying distribution, the expectation values of the squares of these terms satisfies
\begin{equation*}
\sum_{\mathbf{j}\neq 0^{\ell}} \left\langle \left( v_{\mathbf{j}}^{(1)} \frac{1}{\sqrt{2^{\ell(r-1)}}}\right)^2\right\rangle = \frac{1}{2^{\ell(r-1)}}\left( 1-\frac{1}{2^{\ell}}\right)
\end{equation*}
For each value of $\mathbf{j}$, this is a diagonal term of $\chi$ and $\mathbf{j}$ takes $4^{\ell}-1$ values in the sum. Therefore, there are $4^{\ell}-1$ diagonal elements of $\chi$ that sum upto the above expression, regardless of the distribution.  Thus, at least one of these diagonal terms must be larger than their average,  and the largest singular value is larger than all diagonal elements of $\chi$. Thus, 
\begin{equation}
\sigma_{\max}(\chi)\geq \frac{1}{4^{\ell}-1}\frac{1}{2^{\ell(r-1)}}\left( 1-\frac{1}{2^{\ell}}\right) = \frac{1}{2^{N-m}(2^{\ell}+1)}
\end{equation}
Here, we have used $r\ell = N-m$ $\blacksquare$

Thus, it is impossible to improve the convergence rate without using entanglement.  In the next section, we will develop a physical intuition to understand this fact. 

\tocless\subsection{Physical interpretation of the proofs of the convergence theorems}\label{intuition}

 Although the proofs of the two theorems appear technical, there is an insightful physical picture to interpret the singular values of $\chi$, connecting it back to the Hierarchy of subsystems introduced in Fig.~\ref{FIG1}a.  They can be understood as the information contained in various levels of the subsystem Hierarchy. Let us consider a quantum state $\rho$ of $N-m$ qubits.  We can write it as  $\rho =\sum_{\mathbf{i}}v_{\mathbf{i}}\sigma_{\mathbf{i}}$. The indices $\mathbf{i}$ fit well within the Hierarchy of subsystems. For each index, one can define a $s(\mathbf{i})=$ subset of qubits corresponding to which $ \mathbf{i}$  has a non-zero character.  For example, $s(000\cdots 0)=\emptyset$, $k(21000\cdots 0)=\{1, 2\}$.  Thus, every index $\mathbf{i}$ and every matrix $\sigma_{\mathbf{i}}$ belongs to a specific level in the Hierarchy of subsystems.  We can now define the weight of $\rho$ in each of these hierarchical levels.  We define 
\begin{equation*}
w_{s} = \frac{1}{3^k} \sum_{s(\mathbf{i})=s} v_{\mathbf{i}}^2
\end{equation*}
Here, $k$ is the number of qubits in $s$. Note that we have normalized the overlap by the dimension given by $3^k$.  $w_s$ represents the part of the information in $\rho$ contained as the correlation between qubits in $s$ of the Hierarchy. For instance, if we consider a partial trace of $\rho$ with less than $k$ qubits remaining, then we lose all of the $v_{\mathbf{i}}$'s contributing to e$w_s$. In other words, we lose the information carried by the  higher levels in the Hierarchy.  To consider an extreme example, if $\rho = \mathbbm{1}/2^{N-m}$, then $w_s=0$  for every non-trivial $s$ --- \textit{no information} is contained in any order higher than zero.  If $\rho$ is a pure state, then it follows that $\sum 3^k w_s =1$ and $w_{\emptyset}=\frac{1}{2^{N-m}}$. Thus, there is necessarily information stored in higher levels.  Since states of quantum systems in a single realization is always pure, the challenge is to ease the information in higher orders, which, according to the above two theorems, can be done  efficiently only in the presence of entanglement.

To understand the role of entanglement, let us consider the average value of $w_s$ over Haar random states and over separable states.  These average values of $w_s$ are precisely the diagonal entries of  $\chi$ and moreover, the off-diagonal entries of $\chi$ are zero, making $w_s$ the singular values of $\chi$.  Intuitively, the  convergence rate to the uniform mixed state is limited by the largest of these $w_s$, because for a uniform mixed state, each $w_s$ is zero except for $s=\emptyset$.  Fig.~\ref{FIG3}b shows $w_s$ for Haar random states (black) and for product states (grey) for $N-m$. Fig.~\ref{FIG3}c shows the same for $N-m=9$. While both of them satisfy the sum condition,  $\sum 3^k w_s =1$, the Haar random states have a uniform distribution of information across the levels of Hierarchy, making the convergence rate fastest possible.  Therefore, Haar random states are highly entangled states and have information \textit{scrambled} uniformly across the Hierarchy, which allows for a faster convergence.  Experimentally, however, producing a Haar random $\ell-$ bit state is much harder than producing \textit{one specific} $\ell-$bit entangled state.  One can consider a protocol where a sample state of $\bar{S}$ is produced by applying independent random single qubit gates on a \textit{specific} $\ell-$bit entangled state.  This is a more realistic protocol, experimentally.  A natural question arises: Which specific $\ell-$bit entangled states, upon averaging over single qubit rotations, produces the uniform mixed state at a faster convergence rate? 

In quantum metrology, one of the celebrated examples of quantum enhancement is using a GHZ state~\cite{PhysRevLett.96.010401, RevModPhys.90.035005}. In a curious analogy, we find that using a generalization of GHZ states, known as $k-$uniform states~\cite{PhysRevA.90.022316}, one can enhance the convergence rate of the errors in preparing the uniform mixed state. \\

\tocless\section{Quantum speed-up in preparation}\label{speed_up_1}
In this section, we will show two classes of specific $\ell-$ bit states which, upon applying only single-qubit random gates,  converge rapidly to the uniform mixed states. Unlike Haar-random $\ell-$bit states where we need a reliable way of preparing \textit{all} $\ell-$bit states, the results in this section can be implemented experimentally with a reliable preparation of only one specific $\ell-$bit state.  We first begin with a theorem that helps us compute the convergence rate for the case when we start with a fixed state followed by random single qubit gates.

\begin{figure}
\includegraphics[scale=1]{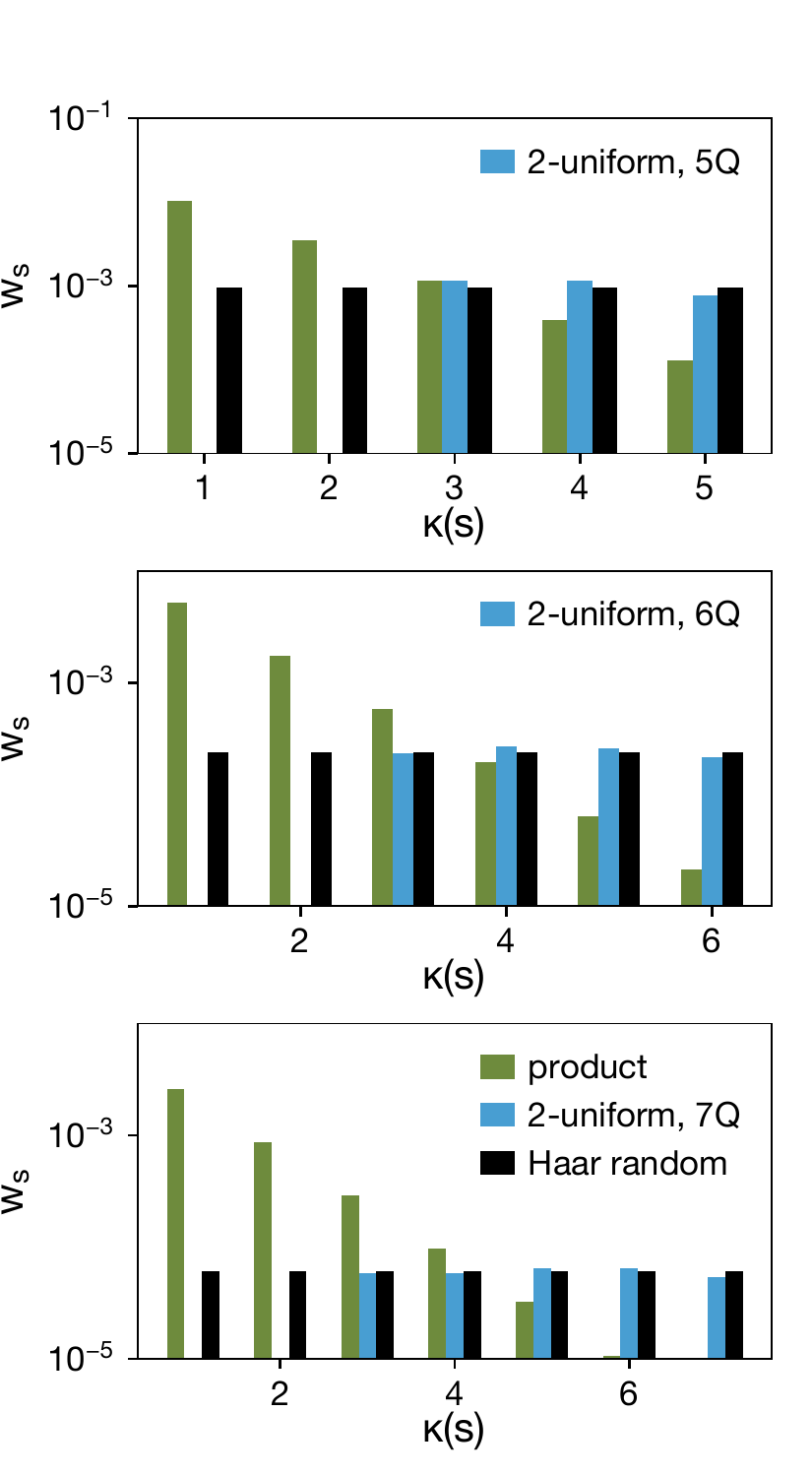}
\caption{\textbf{$k_0-$uniform states: }The distribution of $w_s$ across the Hierarchy for elementary $2-$uniform states with $5, 6$ and $7$ qubits, listed in  ref.~\cite{PhysRevA.90.022316}, compared with the corresponding distributions for Haar random states and product states.  The convergence rate, while using these $2-$uniform states is almost as fast as Haar-random states. For a general result on the convergence rate, see 
convergence theorem $4$ in the text.}\label{FIG4}
\end{figure}

\textbf{Convergence theorem - 3}: If we use $\nu$ experimental shots starting from a fixed  state $\rho_{\bar{S}}$ for the qubits in $\bar{S}$, followed by an independent random single qubit rotation on each qubit to produce the state $\rho_{\text{real}, \bar{S}}$, the $\chi-$matrix of the resulting error $\epsilon = \frac{1}{2^{N-m}}\mathbbm{1}_{\bar{S}}-\rho_{\text{real}, \bar{S}}$ satisfies 
\begin{equation}
\sigma_{\max}(\chi) = \frac{1}{\nu}\max_s \{w_s\}
\end{equation}
Here, $w_s$ are the weights of fixed state $\rho_{\bar{S}}$ corresponding to the subset $s\subset \bar{S}$.  

Moreover, if $\rho$ is a fixed $\ell-$bit state with weights $\{w_k\}$ and we start with the state $\rho\otimes \cdots \otimes \rho$ of the $N-m$ qubits in $\bar{S}$, followed by independent random single qubit rotations, the resulting error is
$$
\sigma_{\max}(\chi) = \frac{1}{\nu 2^{N-m-\ell}}\max_s \{w_s\}
$$

\noindent \textbf{Proof:} For a state $\rho =\sum_{\mathbf{i}}v_{\mathbf{i}}\sigma_{\mathbf{i}}$,  it follows by definition, in the basis $\{\sigma_{\mathbf{i}}\}$ that
$$
\chi_{\mathbf{i};\mathbf{j}} = \langle v_{\mathbf{i}}v_{\mathbf{j}}\rangle
$$
when $\mathbf{i}, \mathbf{j}\neq 00\cdots 0$.  The averaging $\langle \cdots \rangle$ is over single qubit rotations.  It follows that $\langle v_{\mathbf{i}}v_{\mathbf{j}}\rangle=0$ when $\mathbf{i}\neq \mathbf{j}$.  Moreover, the only invariant of a vector under $SO(3)$ is it's length. Therefore, it follows that $\langle v_{\mathbf{i}}^2\rangle=w_{s(\mathbf{i})}$. Thus, the singular values of $\chi$ are $w_s$ for $s\neq \emptyset$. The theorem now follows $\blacksquare$

This theorem shows that $\max_s\{w_s\}$ of the initial state completely determines the convergence rate. Therefore,  order to find the  fastest converging protocol to prepare the qubits in $\bar{S}$ in the uniform mixed state by applying random single qubit gates on a fixed entangled initial state, it suffices to optimize $\max_s\{w_s\}$ of the initial state. \\

\begin{figure}
\includegraphics[scale=0.9]{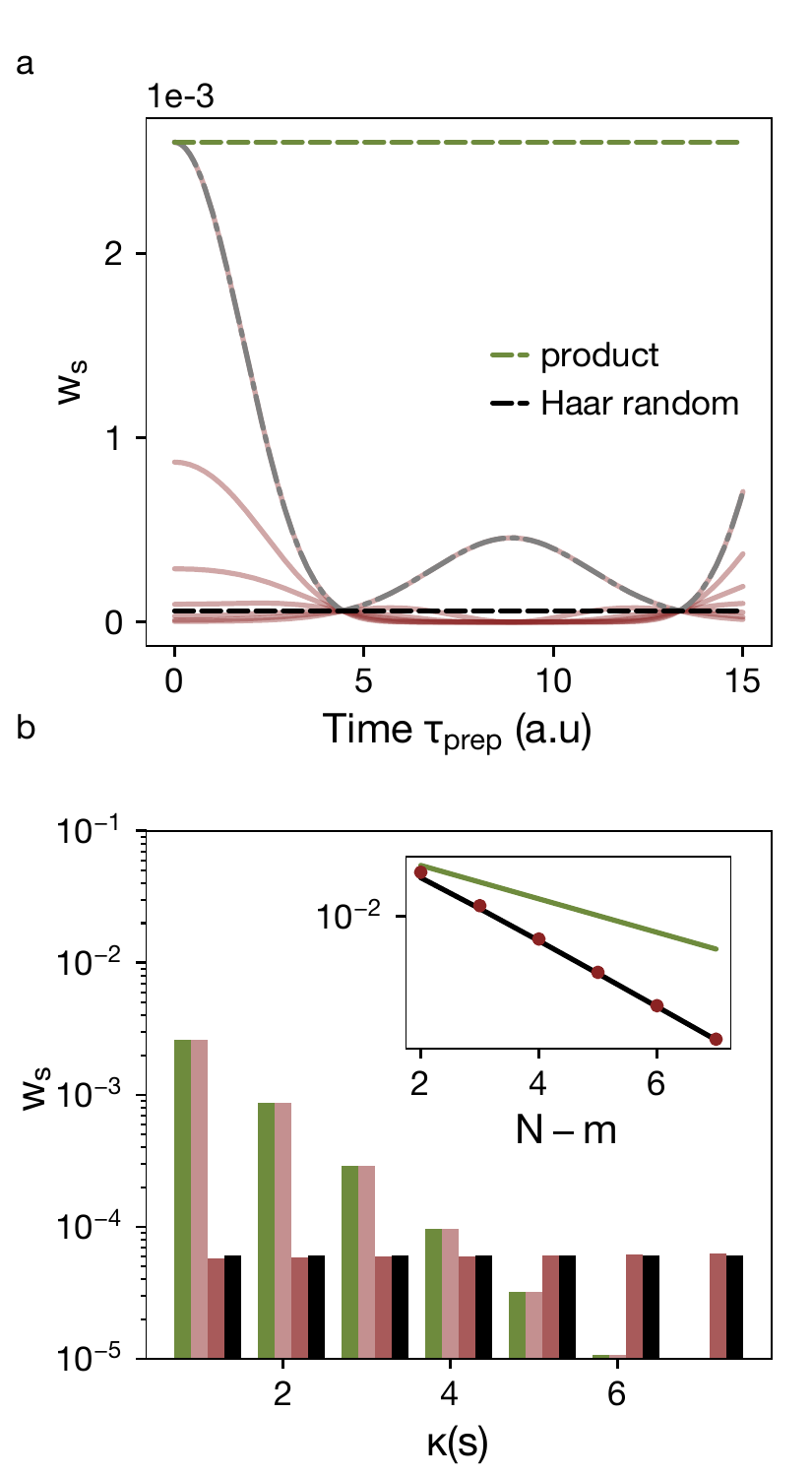}[h]
\caption{\textbf{Information scrambling: }\textbf{a} $w_s$ for a few choices of the subset $s$ of $N-m=7$ qubits as a function of time, evolving under the Heisenberg chain $H_{\text{prep}}$.  The gray line shows the maximum $w_s$. The green and black dashed lines represent the limits of product and Haar-random states respectively.  One can see that at specific times, the maximum $w_s$ is close to that of a Haar-random state.  These calculations were performed with $J=h=1$.   \textbf{b} shows a distribution of $w_s$ for the initial state $\ket{\psi(0)}$ (light shade of brown) and $\ket{\psi(\tau_{\text{prep}})}$ (dark shade of brown) at the optimal time $\tau_{\text{prep}}$, compared with Haar-random and product states. The Inset shows the scaling of the convergence rates with $N-m$. The brown points represent the scaling of $\max_s \{w_s\}$ for $\ket{\psi(\tau_{\text{prep}})}$, and it is very close to a Haar random state.}\label{FIG5}
\end{figure}

\tocless\subsection{ $k$-Uniform states}
The insights developed in Sec.~\ref{intuition} suggest that the key to fast convergence is scrambling of information across the various levels of the Hierarchy.  Moreover, Fig.~\ref{FIG3}c shows that the problem with product states is, the weights $w_s$ for small subsets $s$ are too high. We therefore consider the $\ell-$bit states $\rho$ where the weights $w_s$ are zero for small subsets. That is, we consider pure states $\rho$ where $w_s=0$ for $\kappa(s)\leq k_0$, for subsets with  $k_0$ or fewer qubits.  Such states have the property that their partial trace corresponding to any subset containing $k_0$ or fewer qubits is equal to the uniform mixed state. This is clear from the Hierarchical picture Fig.~\ref{FIG1}a.  These are known as \textit{$k_0-$uniform states}~\cite{PhysRevA.90.022316}.  They are generalizations of the GHZ state. Note that the GHZ state is $1-$uniform.  It follows that $k_0\leq \lfloor{\ell/2}\rfloor$.  It is easy to see that such states improve the convergence rate.

\textbf{Convergence theorem - 4}: If we use $\nu$ experimental shots with a fixed $\ell-$bit, $k_0-$uniform state for the qubits in $\bar{S}$, followed by a random single qubit rotation on each qubit to produce the state $\rho_{\text{real}, \bar{S}}$, the $\chi-$matrix of the resulting error $\epsilon = \frac{1}{2^{N-m}}\mathbbm{1}_{\bar{S}}-\rho_{\text{real}, \bar{S}}$ satisfies 

\begin{equation}
\sigma_{\max}(\chi)\leq \frac{2^{k_0+1}-1}{\nu 2^{N-m}3^{k_0+1}}
\end{equation}

\noindent \textbf{Proof:} Let $\rho =\sum_{\mathbf{i}}v_{\mathbf{i}}\sigma_{\mathbf{i}}$ be an $N-m$ qubit state of the qubits in $\bar{S}$.  Let $s\subset \bar{S}$ with $k$ qubits.  We consider the partial trace of $\rho$ corresponding to  the qubits in $s$.  $\text{Tr}_{\bar{S}-s}\rho =\sum{\mathbf{i}}u_{\mathbf{i}}\sigma_{\mathbf{i}}$. We will now consider the final terms in this sum, i.e., terms with $\mathbf{i}$ containing no zeros. There are $3^k$ such terms.  It follows from  $\sum{\mathbf{i}}u_{\mathbf{i}}^2\leq 1 $ that $\sum_{k(\mathbf{i})=k} u_{\mathbf{i}}^2\leq 1-u_{00\cdots 0}^2 =1-1/2^k$ for each $\mathbf{i}\neq 00\cdots 0$. Finally, by definition of the partial trace, for $\kappa(\mathbf{i})=k$, $u_{\mathbf{i}}=\frac{1}{\sqrt{2^{N-m-k}}}v_{\mathbf{i}}$. Thus,
\begin{equation*}
\begin{split}
w_s& = \frac{1}{3^k}\sum_{\mathbf{i}\sim s}v_{\mathbf{i}}^2 \\
&= \frac{1}{3^k}\frac{1}{2^{N-m-k}}\sum_{\kappa(\mathbf{i})=k}u_{\mathbf{i}}^2 \leq \frac{1}{3^k} \frac{1}{2^{N-m-k}}\left(1-\frac{1}{2^k}\right)
\end{split}
\end{equation*}

Note that if $\rho$ is $k_0$ uniform, $w_s=0$ unless $k\geq k_0+1$. Therefore using convergence theorem -3,  it follows that 
$$
\sigma_{\max}(\chi) \leq \frac{2^{k}-1}{\nu 2^{N-m}3^{k}}
$$
for each $k > k_0$.  Noting that the above is a decreasing function in $k$, the result follows. $\blacksquare$

In Fig.~\ref{FIG4} we show some exact calculations of $w_s$ for elementary $2-$uniform states with $5$, $6$ and $7$ qubits listed in ref.~\cite{PhysRevA.90.022316}. While the above theorem shows an exponential advantage for $k_0-$uniform states, constructing such states for large $k_0$ is a largely open problem even theoretically and experimentally expected to be very challenging. Therefore, we propose an alternative experimentally implementable initial states based on information scrambling, taking a cue from the recent works on thermalization and localization~\cite{scherg_observing_2020, HS_fragmentation}.

\tocless\subsection{Information scrambling}
In this section, we will present a more realistic experimental protocol to speed up the preparation of $\mathbbm{1}_{\bar{S}}$.  Information scrambling and entanglement scrambling~\cite{scrambling} have been studied in several many-body systems.  We study the following protocol: (i) initiate the $N-m$ qubits in a product state. (ii) evolve the system under a many-body Hamiltonian $H_{\text{prep}}$, which is known to scramble the entanglement, e.g. Heisenberg model, for a fixed duration $\tau_{\text{prep}}$ and (iii) apply random single qubit gates on each qubit.  If the distribution of $w_s$ of the state prepared by the evolution under $H_{\text{prep}}$ is sufficiently uniform, this protocol will result in a speed up of the convergence rate of the state of the qubits to the uniform mixed state.

We choose the simple Heisenberg chain:
\begin{equation}
H_{\text{prep}} = J\sum_i \sigma_{x, i}\sigma_{x, i+1} + \sigma_{yi}\sigma_{y, i+1} + \sigma_{z,i}\sigma_{z,i+1} + h\sum_i \sigma_{z, i}
\end{equation}

We start with an initial state $\ket{\psi(0)} = \frac{1}{2^{\frac{N-m}{2}}} $ and after the evolution under the above Hamiltonian, the state is $\ket{\psi(\tau_{\text{prep}})} =e^{-i H_{\text{prep}}\tau_{\text{prep}} }\ket{\psi(0)} $. One can clearly see in Fig.~\ref{FIG5}a that at particular times, the $\max_s \{w_s\}$ is close to the lower limit shown in convergence theorem 1. Fig.~\ref{FIG5} shows the distribution of $w_s$ for this state with $N-m=7$ at the particular time.  Finally the inset in Fig.~\ref{FIG5}b shows  that the scaling of $\max_s \{w_s\}$ at the optimal time is indeed the same as that for a Haar random state. 

Thus, one can achieve an exponential speed-up in the convergence rate in state of the art experiments using the above method. However, scrambling information all across $N-m$ qubits may be hard with large numbers of qubits. In those cases, we can group the $N-m$ qubits into $\ell-$bits and scramble the entanglement within each group. This will provide a convergence rate for the state preparation sampling error of $\frac{1}{\sqrt{\nu}\sqrt{2^{\ell}+1}}$, which even for modest values of $\ell =10$ provide more than two orders of magnitude improvement in the demand for $\nu$.  \\

\tocless\section{Quantum speed up in measurement}\label{speed_up_2}
We will now address the rate of convergence of the readout sampling error $\Delta^2_{\text{meas.}}\Phi_S$ and present metrological techniques to enhance the convergence rate.  Recently there have been a few investigations into this  and related problems~\cite{https://doi.org/10.48550/arxiv.2210.03030, https://doi.org/10.48550/arxiv.2209.14328}. We begin with a reformulation of Fisher information and Cramer-Rao bound in the language of the Choi matrix, as most metrological advantages are established using this approach~\cite{PhysRevLett.96.010401}. 

\tocless\subsection{Fisher Information: General Framework}
Let $\rho_{f}=\Phi_S(\rho_{i})$ be the final state after the operation $\Phi$ is applied on $\rho_i$.  If $\Phi$ is characterized by parameters $\{\phi_1, \cdots, \phi_r\}$,  the quantum metrological problem is to find the optimal initial state $\rho_i$ and a measurement so as to estimate these parameters.  The relevant quantity is the fisher information  defined as,
\begin{equation}
F_{\alpha \beta}=\frac{1}{2}\text{Tr}[ \{L_{\alpha}, L_{\beta}\} \rho]
\end{equation}
The diagonal element  $F_{\alpha \alpha}$ represent the Fisher information corresponding to the variable  $\phi_{\alpha}$.  Here $L_{\alpha}$ is the quantized logarithmic derivative of $\rho$ w.r.t $\phi_{\alpha}$, defined as
\begin{equation*}
\frac{1}{2}\{L_{\alpha}, \rho \} = \frac{d\rho }{d \phi_{\alpha}}
\end{equation*}
We will develop a general method to compute the Fisher information using \textit{dualities}.  We define $\{\tilde{L}_\mathbf{i}\}$ for $\mathbf{i} \in \{0, 1, 2, 3\}^N$ as duals of the pauli operators $\sigma_{\mathbf{i}}$ defined earlier.  Let us consider a general $\rho = \sum_{\mathbf{i}} v_{\mathbf{i}} \rho_{\mathbf{i}}$.  We define
\begin{equation}
\frac{1}{2}\{\tilde{L}_{\mathbf{i}}, \rho\} = \sigma_{\mathbf{i}}
\end{equation}
We discuss the details of how to compute the duals in the supplementary materials. Moreover, we define
\begin{equation}
\mathcal{F}^{\mathbf{k}}_{\mathbf{i}\mathbf{j}} = \frac{1}{2}\text{Tr}[\{\tilde{L}_{\mathbf{i}}, \tilde{L}_{\mathbf{j}}\}\sigma_{\mathbf{k}}]
\end{equation}
We can now use this tensor to construct the relevant Fisher information, using the observation $\partial_{\alpha} = \sum_{\mathbf{i}}\frac{d v_{\mathbf{i}}}{ d \alpha} \partial_{\mathbf{i}}$.
\begin{equation}
L_{\alpha} = \sum_{\mathbf{i}}\frac{dv_{\mathbf{i}}}{d\phi_{\alpha}} \tilde{L}_{\mathbf{i}}
\end{equation}
Therefore, the Fisher information can be written as 
\begin{equation}
F_{\alpha\beta} = \partial_{\alpha}v_{\mathbf{i}} \partial_{\beta} v_{\mathbf{j}} \mathcal F_{\mathbf{i}\mathbf{j}}^{\mathbf{k}}v_{\mathbf{k}}
\end{equation}
Representing the unit vector $\mathbf{v}=(v_{00\cdots}, \cdots)$, we may write the Fisher information as 
\begin{equation}
F_{\alpha\beta}  = (\partial_{\alpha} \mathbf{v})^T (\mathcal F^{k}v_k) (\partial_{\beta} \mathbf{v})
\end{equation}
The magnitude of the Fisher information depends on the length of the vectors $\{\partial_{\alpha}\mathbf{v}\}$.  Intuitively, one can understand this in the following way. The relevant information is contained in  the vector $\mathbf{v}(\phi_1, \cdots, \phi_r)$ and therefore, the limits on the precision in estimating a parameter $\phi_{\alpha}$ is related to the \textit{sensitivity} of $\mathbf{v}$ to that parameter, i.e., magnitude of $\partial_{\alpha}\mathbf{v}$. We will illustrate this point using a few examples.

\tocless\subsection{GHZ states}
Let us consider a standard interferometry, under a unitary $U=e^{-i\frac{\phi}{2}\sigma_z}$ ~\cite{PhysRevLett.96.010401}. The corresponding Choi matrix is given by 
\begin{equation}
\begin{split}
\rho^{\Phi} =& \frac{1}{2}[\mathbbm{1} + \cos \phi \sigma_x\otimes \sigma_x+ \cos \phi \sigma_y\otimes \sigma_y\\
&+ \sin \phi \sigma_x\otimes \sigma_y- \sin \phi \sigma_y\otimes \sigma_x]
\end{split}
\end{equation}
If we apply this operation on $\nu$ qubits, the resulting Choi matrix is $(\rho^{\Phi})^{\otimes \nu}$.  We consider two initial states. First, a separable state $(\rho_i)^{\otimes \nu}$ with $\rho_i=\frac{1}{2}(\mathbbm{1}+\sigma_x)$. The final state is $(\rho_f)^{\otimes \nu}$ with 
$$\rho_f = \frac{1}{2}(\mathbbm{1}+\cos \phi \sigma_x + \sin\phi \sigma_y) = \frac{1}{\sqrt{2}}\sigma_0 + \frac{\cos\phi}{\sqrt{2}}\sigma_1 + \frac{\sin\phi}{\sqrt{2}}\sigma_2$$
Therefore, the relevant vector is $\mathbf{v} = (1/\sqrt{2}, \cos \phi/\sqrt{2}, \sin\phi/\sqrt{2})^{\otimes \nu}$. It is straightforward to see that 
\begin{equation}
|\partial_{\phi}\mathbf{v}|^2 =\frac{\nu}{2}
\end{equation}.
This is the standard quantum limit. More generally, if $\mathbf{v}=\mathbf{u}_1\otimes \cdots \otimes \mathbf{u}_{\nu}$, and each $\mathbf{u}_{i}$ is a unit vector (i.e., the states are pure), it follows that 
$$
|\partial_{\phi}\mathbf{v}|^2 = \sum_i |\partial_{\phi}\mathbf{u}_i|^2
$$
Let us now consider a GHZ state input, $\ket{GHZ}=\frac{1}{\sqrt{2}}(\ket{0}^{\otimes \nu}+\ket{1}^{\otimes \nu})$. We make a few observations in order to write the corresponding density matrix in the pauli basis.  For an operator, $\bra{GHZ}\sigma_{\mathbf{i}}\ket{GHZ}=0$, unless $\mathbf{i}\in\{0, 3\}^\nu$ or  $\mathbf{i}\in\{1, 2\}^\nu$. That is, if $\mathbf{i}$ contains a $0$ or a $3$ and a $1$ or a $2$, the expectation is zero and this can be seen easily.  Moreover, it follows that for $\mathbf{i}\in\{0, 3\}^\nu$
\begin{equation}
\bra{GHZ}\sigma_{\mathbf{i}}\ket{GHZ}=\begin{cases}
0 & \text{ if } k_z(\mathbf{i}) = \text{ odd }\\
1/\sqrt{2^{\nu}} & \text{ if } k_z(\mathbf{i}) = \text{ even }\\
\end{cases}
\end{equation}
Here, $k_z(\mathbf{i})=$number of $3$'s in $\mathbf{i}$. And for $\mathbf{i}\in \{1, 2\}^\nu$, 
\begin{equation}
\bra{GHZ}\sigma_{\mathbf{i}}\ket{GHZ}=\begin{cases}
0 & \text{ if } k_y(\mathbf{i}) = \text{ odd }\\
(-1)^{k_y(\mathbf{i})/2}/\sqrt{2^{\nu}}  & \text{ if } k_y(\mathbf{i}) = \text{ even }\\
\end{cases}
\end{equation}
Here, $k_y(\mathbf{i})=$number of $2$'s in $\mathbf{i}$. Thus, 
\begin{equation}
\begin{split}
\rho_{GHZ} = &\frac{1}{\sqrt{2^{\nu}}} \sum_{\mathbf{i_z}\in\{0,3\}^\nu, k_z=\text{even}}\sigma_{\mathbf{i_z}} \\
&+  \frac{1}{\sqrt{2^{\nu}}}\sum_{\mathbf{i_{xy}}\in \{1, 2\}^\nu, k_{y}=\text{even}}(-1)^{k_y(\mathbf{i_{xy}})/2}\sigma_{\mathbf{i_{xy}}} 
\end{split}
\end{equation}
Note that the second term is a $\nu-$qubit correlator.  This term can also be written as 
\begin{equation}
\begin{split}
 &\frac{1}{\sqrt{2^{\nu}}}\sum_{\mathbf{i_{xy}}\in \{1, 2\}^\nu, k_{y}=\text{even}}(-1)^{k_y(\mathbf{i_{xy}})/2}\sigma_{\mathbf{i_{xy}}} \\
 &= \frac{1}{2}[(\sigma_1 + \ii \sigma_2)^{\otimes \nu}+(\sigma_1 - \ii \sigma_2)^{\otimes \nu}]
\end{split}
\end{equation}
The GHZ state has the special property that \textit{half} the magnitude of the vector $\mathbf{v}$ is held by the highest order correlators.  That is it has a \textit{heavy tail}, which we will show is responsible for the quantum enhancement in Fisher information.  The final state after the operation $(\rho^{\Phi})^{\otimes \nu}$ is 

\begin{equation}
\begin{split}
\rho_f = &\frac{1}{\sqrt{2^{\nu}}} \sum_{\mathbf{i_z}\in\{0,3\}^\nu, k_z=\text{even}}\sigma_{\mathbf{i_z}} \\
&+  \frac{1}{2}[e^{-i\nu \phi}(\sigma_1 + \ii \sigma_2)^{\otimes \nu}+e^{i\nu \phi}(\sigma_1 - \ii \sigma_2)^{\otimes \nu}]
\end{split}
\end{equation}
It is straightforward to see that 
$$
|\partial_{\phi}\mathbf{v}|^2 = \nu^2/4
$$
This shows the enhanced sensitivity, concurrent with the Fisher information. Therefore, $|\partial_\phi \mathbf{v}|^2$ is a proxy for the Fisher information.  Below, we generalise this protocol for a general reduced process tomography of $1$ qubit. \\

\tocless\subsection{A protocol for the general case}
In this section, we will develop a generalized protocol based on GHZ states to speed-up the convergence of the sampling error $\Delta^2_{\text{meas.}} \Phi_S$.  We consider $m=1$ qubit in $S$ for simplicity.  The reduced Choi matrix $\rho^{\Phi, S}$ is a $4\times 4$ doubly stochastic map, i.e., both the partial traces of this matrix are equal to $\mathbbm{1}$.  This follows from $\text{Tr}\Phi_S(\rho_S)=\text{Tr}(\rho_S)$ and $\Phi_S(\mathbbm{1})=\mathbbm{1} $. Thus, it has $9$ free parameters and it can be written as 
\begin{equation}
\rho^{\Phi, S} = \frac{1}{2}\mathbbm{1} +\sum_{i, j \in \{x, y, z\}} \phi_{ij}\sigma_{i}\otimes \sigma_j
\end{equation}
That is, it contains only terms of the form $\sigma_x\otimes \sigma_x$, $\sigma_x\otimes \sigma_y$ etc. If we start with a GHZ state of $\nu$ qubits and apply this operation on each one of the qubits, the final state is
\begin{equation}
\begin{split}
\rho_f = &\frac{1}{\sqrt{2^{\nu}}} \sum_{\mathbf{i_z}\in\{0,3\}^\nu, k_z=\text{even}} \phi_{zz}^{k_z(\mathbf{i_z})}\sigma_{\mathbf{i_z}} \\
&+  \frac{1}{2}\left[((\phi_{xx}+i\phi_{yx})\sigma_1 + (\phi_{xy}+i\phi_{yy}) \sigma_2)^{\otimes \nu}\right.\\ &+\left. ((\phi_{xx}-i\phi_{yx})\sigma_1 + (\phi_{xy}-i\phi_{yy})\sigma_2)^{\otimes \nu}\right]
\end{split}
\end{equation}
Note that this state is insensitive to $\phi_{xz}, \phi_{zx}, \phi_{yz}$ and $\phi_{zy}$.  It follows from the previous discussion that the Fisher information corresponding to the relative phase  between $\phi_{xx}+i\phi_{yx}$ and $\phi_{xy}+i\phi_{yy}$ scales as $\nu^2$ and this parameter can be measured efficiently. By using $SU(2)\times SU(2)$ operations on $\rho^{\Phi, S} $, we can construct similar protocols to measure $8$ such phases, all independent. Thus, one can use this protocol to efficiently estimate $8$ out of the $9$ parameters $\phi_{ij}$.  The $9$-th parameter is $\sum_{ij}\phi_{ij}^2$, the \textit{purity} of $\rho^{S, \Phi}$, which is invariant under $SU(2)\times SU(2)$~\cite{PhysRevA.89.062110}.  The measurement of this quantity cannot be made efficient using the above GHZ state technique.  This quantity is called the \textit{spreading parameter} of a noisy channel and it can be estimated efficiently using squeezed states, as has been shown recently~\cite{https://doi.org/10.48550/arxiv.2208.09386}. GHZ states have been demonstrated in Rydberg atoms trapped in a tweezer array~\cite{Omran_2019} and can be used to implement the above outlined protocol.  Combining this protocol with the one outlined in Sec.~\ref{speed_up_1}, we obtain a general protocol for metrologically enhanced reduced process tomography (Fig.~\ref{Fig2}).\\

\tocless\section{Conclusion and Outlook}

To summarize, we introduced a new approach towards benchmarking analog and many-qubit quantum operations, based on the reduced Choi matrix. A tomography of the reduced Choi matrix is met with two courses of sampling errors --- one at the preparation of mixed initial states which are necessary for a reduced process tomography and the other at the measurement stage.  We showed fundamental limits to the convergence rate of the sampling error in state preparation and established an analogy between these limits and the standard quantum limit and Heisenberg limits in quantum metrology. Moreover, we developed an efficient protocol to produce these mixed states. We also developed quantum metrology protocols to optimize the convergence rate of the sampling error in the measurements, based on variants of existing protocols using GHZ states.  We address the computational aspects of benchmarking --- i.e., the question of how to obtain benchmarks using a reduced Choi matrix in the next part of this paper. We will show that for an analog quantum operation, symmetries of the applied Hamiltonian can be utilised to develop benchmarks. More generally, the reduced Choi matrix corresponding to a unitary map is always doubly stochastic~\cite{landau1993birkhoff, GOWDA201740}.  In the next paper, we will show that violation of double stochasticity can be used to develop benchmarks of the quantum operation, which also naturally leads to an error mitigation technique. 

The GHZ--state based protocol for reduced process tomography  introduced in this paper can be developed further, to develop protocols using other, more experimentally accessible states. We have shown that the key property of GHZ states that lead to metrological advantage is that it has a \textit{heavy tail}. One can construct other states with the same property and develop other metrology protocols for reduced process tomography.  The protocols already presented in the paper can be readily implemented in state-of-the art experimental platforms including Rydberg atoms trapped in a tweezer array.

\paragraph*{\textbf{Acknowledgments}}
I thank Monika Aidelsburger for illuminating ``hallway" discussions.  This work was supported by the European Union and Deutsche Forschungsgemeinschaft (DFG, German Research Foundation) under Germany's Excellence Strategy -- EXC- 2111 -- 39081486. The work at LMU was additionally supported by DIP . This project has received funding from the European Union’s Horizon 2020 research and innovation programme under the Marie Skłodowska-Curie grant agreement No 893181.

\paragraph*{\textbf{Competing interests}} The authors declare no competing interests. 

\bibliography{References}


\appendix

\cleardoublepage

\setcounter{figure}{0}
\setcounter{page}{1}
\setcounter{equation}{0}
\setcounter{section}{0}

\renewcommand{\thepage}{S\arabic{page}}
\renewcommand{\thesection}{S\arabic{section}}
\renewcommand{\theequation}{S\arabic{equation}}
\renewcommand{\thefigure}{S\arabic{figure}}
\onecolumngrid
\begin{center}
\huge{Supplementary Information}
\vspace{5mm}
\end{center}
\twocolumngrid
\normalsize
\tableofcontents

\section{Properties of the Choi matrix}
In this section, we will summarize the properties of the Choi matrix. For a $d$-dimensional quantum system, the states are represented by hermitian, positive semi-definite $d\times d$ matrices with a unit trace. General quantum operations on this system are represented by Choi matrices, which are $d^2\times d^2$, Hermitian, positive semi-definite matrices. If the map is trace preserving, the Choi matrices satify an additional condition --- each $d\times d$ diagonal block has a unit trace.  Thus, the trace of the Choi matrix is $d$.  If $\{\ket{1}, \cdots, \ket{d}\}$ is a basis set and $\Phi$ is a quantum process, its Choi matrix is given by $\rho^{\Phi}_{ij; kl} = \text{Tr}(\Phi(\ket{i}\bra{j})\ket{k}\bra{l})$. 

Let us consider a simple example. If the map $\Phi$ is an identity, i.e., $\Phi: \ket{j}\mapsto \ket{j}$, the Choi matrix is given by $\rho^{\Phi}_{ij; kl} = \delta_{ik}\delta_{jl}$.  For $d=2$, this is
\begin{equation*}
\rho^{\Phi} = \left(
\begin{array}{cccc}
1 & 0 & 0 & 1\\
0 & 0 & 0 & 0\\
 0& 0 & 0 & 0\\
1 & 0 & 0 & 1\\
\end{array}
\right)
\end{equation*}
Indeed, this looks similar to the density matrix of a  Bell state. More generally, the Choi matrix corresponding to the identity map is
\begin{equation}
\rho^{\Phi} =\left( \sum_i  \ket{i}\otimes\ket{i}\right) \left( \sum_i  \bra{i}\otimes\bra{i}\right)
\end{equation}
One can interpret this in the following way. The Choi matrix describes a bipartite quantum state of $d-$ dimensions each. The first $d-$dimensional system can be thought of as the input and the second can be thought of as the output.  The identity map, maps each input state $\ket{i}$ to the output $\ket{i}$. Thus, the process is a superposition of the products $\ket{i}\otimes \ket{i}$.  As a second example, let us consider a unitary operation $\Phi: \ket{j}\mapsto U\ket{j}$. The Choi matrix is
\begin{equation*}
\rho^{\Phi} = \left( \sum_i  \ket{i}\otimes U \ket{i}\right) \left( \sum_i  \bra{i}\otimes\bra{i}U^{\dagger}\right)
\end{equation*}
Note that in both the examples,  the Choi matrix has rank-1, that is it represents a pure state. This however, is not true in general.  For instance, if the process involves a fluctuation of the unitary, then resulting Choi matrix is an average of the Choi matrices corresponding to each unitary.  For a fully decohering quantum process that maps every initial state to the uniform mixed state, that is, $\Phi:\ket{i}\mapsto \frac{1}{d}\mathbbm{1}_d $, the Choi matrix is
\begin{equation*}
\rho^{\Phi} = \frac{1}{d}\mathbbm{1}_{d^2}
\end{equation*}
We can define a process fidelity by $F(\rho^{\Phi}, \rho^U)$.  This is related to the average fidelity by~\cite{Nielsen_2002} 
\begin{equation}
F(\rho^{\Phi}, \rho^U) = \frac{2^N f_{\text{avg}}+1}{2^N+1}
\end{equation}
Both quantum states and quantum processes are described by positive semi-definite matrices. In fact, a quantum process on a $d-$dimensional system can be considered as a quantum state in $d^2$ dimensions. But the converse is not necessarily true. In the following, we will describe some of the similarities and dissimilarities between states and processes. 

\subsection{Similarities between quantum states and quantum processes}
Corresponding to a process represented by a Choi matrix $\rho^{\Phi}$, we can define a state $\frac{1}{d}\rho^{\Phi}$. As we have shown above,  if $\frac{1}{d}\rho^{\Phi}$ as a state is pure, then $\Phi$ is a unitary map.  If $\frac{1}{d}\rho^{\Phi}$ represents a GHZ state, then $\Phi$ is the identity map. If $\frac{1}{d}\rho^{\Phi} = \frac{1}{d^2}\mathbbm{1}$, then $\Phi$ is completely decohering process.  

Beyond these, there is an important result known as the Choi-Jamilkowski isomorphism, which says every trace-preserving Choi matrix can be written as
\begin{equation}
\rho^{\Phi} (\rho) = \sum_{l=1}^{d^2} A_l  \rho A_l^{\dagger}
\end{equation}
with $\sum A_lA_l^{\dagger}= \mathbbm{1}$. One can compare this with the result for mixed states:
\begin{equation}
\rho =\sum \lambda_i \ket{\psi_i}\bra{\psi_i}
\end{equation}
with $\sum \lambda_i =1$. 
Further, every mixed state $\rho$ can be dilated into a pure state. That is, there exists a pure state $\psi \in \mathbbm{C}^d\otimes \mathbbm{C}^d$ such that $\rho$ is the partial trace of $\ket{\psi}\bra{\psi}$. The \textit{Steinspring dilation theorem} shows an analogus result for Choi matrices. Every Choi matrix $\rho^{\Phi}$ can be written as:
\begin{equation}
\Phi(\rho) = \text{Tr}_2 U \rho \otimes \ket{\psi}\bra{\psi} U^{\dagger}
\end{equation}
That is, every quantum process is a partial trace of a unitary map acting on a higher dimensional space.

\subsection{Dissimilarities between quantum states and quantum processes}

The most significant difference between quantum states and Choi matrices is that while \textit{every} positive semi-definite $d^2\times d^2$ matrix with unit trace represents a quantum state,  this is not true for quantum processes. There can be $d^2\times d^2$ postive semi-definite matrices that donot represent any quantum process. For instance, consider $\ket{1}\bra{1}\otimes \ket{1}\bra{1}$. As a state, it represents $\ket{1}\otimes \ket{1}$, while it is not a legitimate quantum process, because it would map $\ket{2}$ to zero.

\section{Moments of the Haar measure}\label{moments}
In this section, we will derive general formulae for the moments of a Haar measure on the space of pure quantum states. We have used some of these formulae in the main text in section III and in the supplementary information  below.  The space of normalized pure quantum states of a $d$-dimensional quantum system is a sphere $\mathbbm{S}^{2d-1}$ in $2d$ dimensional real space: $\mathbbm{S}^{2d-1}=\{(z_1, \cdots, z_d):\ z_i =x_i+\ii y_i, \sum_i x_i^2 +y_i^2=1\}$. For a single qubit, $d=2$ and we obtain $\mathbbm{S}^3$, which foliates into an $ \mathbbm{S}^1-$bundle over the Bloch sphere $\mathbbm{S}^2$ under the Hopf map.  We will compute the moments of the form $\langle |z_{i_1}|^{2r_1}\cdots |z_{i_l}|^{2r_l}\rangle $ under the Haar measure, i.e., the measure invariant under unitary rotations in  $U(d)$.  Here, $r_1, \cdots, r_l$ are integers and $i_1, \cdots, i_l$ are distinct. We make use of the invariance.  

Using the inversion symmetry $z_i \rightarrow -z_i$, it follows that $\langle z_i\rangle =0$.   Moreover, any moment which includes an odd power of a $z_i$ is zero. Therefore, we focus on moments of the form $|z_{i_1}|^{2r_1}\cdots |z_{i_l}|^{2r_l}$. Using $\sum_i |z_i|^2=1$,  and the symmetry, it follows that $\langle |z_i|^2\rangle =\frac{1}{d}$.  In general, we define
\begin{equation}
C_{r_1, \cdots, r_l} = \langle |z_{i_1}|^{2r_1}\cdots |z_{i_l}|^{2r_l}\rangle
\end{equation}
for distinct $i_1, \c dots, i_l$. Note that the above does not depend on the choice of $i_1, \cdots, i_l$.  Let $a=(a_1, \cdots, a_d)$ be a unit vector in $\mathbbm{C}^d$ with length. The projection of any state $\psi =(z_1, \cdots, z_d)$ on $a$, given by $\langle a, \psi\rangle = \sum a_i^* z_i$ has the same mean value as any coordinate $z_i$, which are indeed also projections of the state $\psi$.  That is, 
\begin{equation}
\langle |(a_1^*z_1 +\cdots + a_d^*z_d)|^{2k}\rangle = \langle |z_i|^{2k}\rangle = C_k
\end{equation}
More generally, if $a$ is not a unit vector, it follows that 
\begin{equation}
\langle |(a_1^*z_1 +\cdots + a_d^*z_d)|^{2k}\rangle = |a|^{2k}\langle |z_i|^{2k}\rangle = |a|^{2k}C_k
\end{equation}

Our goal is to evaluate $C_{r_1, \cdots, r_l} $. Let $k=r_1+\cdots+r_l$. It follows that 
\begin{equation}
|(a_1^*z_1 +\cdots + a_d^*z_d)|^{2k} = (a_1^*z_1 +\cdots + a_d^*z_d)^k (a_1z_1^* +\cdots + a_dz_d^*)^k
\end{equation}
Using the multinomial expansion, 
\begin{equation}
\begin{split}
|\sum a_i^*z_i|^{2k} &= \sum_{r_1+\cdots +r_d=k}\Pi_i (a_i^* z_i)^{r_i} \frac{k!}{r_1!\cdots r_d!}\\
& \times \sum_{s_1+\cdots +s_d=k}\Pi_i (a_i z_i^*)^{s_i} \frac{k!}{s_1!\cdots s_d!}
\end{split}
\end{equation}
After averaging over the Haar measure
\begin{equation}
C_k = \sum_{r_1+\cdots +r_d=k}  \Pi_i |a_i|^{2r_i}C_{r_1, \cdots, r_d}\left(  \frac{k!}{r_1!\cdots r_d!}\right)^2
\end{equation}
Furthermore,  the vector $a$ was chosen to be a unit vector. Thus
\begin{equation}
1=\left(\sum_i |a_i|^2\right)^k = \sum_{r_1+\cdots +r_d=k}  \Pi_i |a_i|^{2r_i}  \frac{k!}{r_1!\cdots r_d!}
\end{equation}
 Using the ebove two equations, we obtain
 \begin{equation}
 \begin{split}
 \sum_{r_1+\cdots +r_d=k} & \Pi_i |a_i|^{2r_i} \\
  \times & \left( C_{r_1, \cdots, r_d}\left(  \frac{k!}{r_1!\cdots r_d!}\right)^2 - C_k  \frac{k!}{r_1!\cdots r_d!} \right) = 0
  \end{split}
 \end{equation}
 This must hold for \textit{all} $a$, which implies each term in the above is zero.  Thus,
 \begin{equation}
 C_{r_1, \cdots, r_d} = \frac{r_1!\cdots r_d!}{k!} C_k
 \end{equation}
 
 We can now use the above relation to evaluate $C_k$ and subsequently $C_{r_1, \cdots, r_d} $. We consider
 \begin{equation*}
 1=\left(\sum_i |z_i|^2\right)^k = \sum_{r_1+\cdots +r_d=k}  \Pi_i |z_i|^{2r_i}  \frac{k!}{r_1!\cdots r_d!}
 \end{equation*}
 Evaluating an average over the Haar measure on both sides,
  \begin{equation*}
 1= \sum_{r_1+\cdots +r_d=k}  C_{r_1, \cdots, r_d}  \frac{k!}{r_1!\cdots r_d!} = \sum_{r_1+\cdots +r_d=k} C_k 
 \end{equation*}
 Thus,
 \begin{equation}
 C_k = \frac{1}{\sum_{r_1+\cdots +r_d=k} 1} = \frac{k! (d-1)!}{(d+k-1)!}
 \end{equation}
 And
  \begin{equation}
 C_{r_1, \cdots, r_d} = \frac{r_1! \cdots r_d! (d-1)!}{(d+k-1)!}
 \end{equation}

We list a few particular values
\begin{table}[h!]

\begin{tabular}{|c|c|c|}
\hline
$k$  & Moment & Value \\
\hline 
$k=1$ & $\langle |z_i|^2\rangle$ & $\frac{1}{d}$\\
\hline
$k=2$ & $\langle |z_i|^4\rangle$ & $\frac{2}{d(d+1)}$\\
\hline
$k=2$ & $\langle |z_i|^2 |z_j|^2\rangle$ & $\frac{1}{d(d+1)}$\\
\hline
\end{tabular}

\end{table}

One can derive a similar formula for real vectors $(x_1, \cdots, x_d)\in \mathbbm{S}^{d-1}\subset \mathbbm{R}^d$, using a similar method. Since we use this formula for the real case as well, we briefly describe the derivation.  Under the uniform measure invariant under $SO(d)$, it follows that $\langle x_i^{2k+1} \rangle=0$. Moreover, using $\sum_i x_i^2=1$, it follows that $\langle x_i^2\rangle =\frac{1}{d}$.  Like before, we define
\begin{equation}
C_{r_1, \cdots, r_d}=\langle x_1^{2r_1}\cdots x_d^{2r_d}\rangle 
\end{equation}
Like before, we consider a unit vector $a\in \mathbbm{S}^{d-1}$. It follows that 
\begin{equation}
C_k = \langle (a\cdot x)^{2k}\rangle = \langle (a_1 x_1 + \cdots + a_dx_d)^{2k}
\end{equation}
Using the multinomial theorem,
\begin{equation*}
 (a\cdot x)^{2k} =\sum_{r_1+\cdots+r_d=2k} \Pi_i (a_ix_i)^{r_i}  \frac{2k!}{r_1!\cdots r_d!} 
\end{equation*}
The average vanishes for any odd poers. That is, $\langle x_i^{2k+1} \rangle=0$. Thus,
\begin{equation*}
C_k=\sum_{r_1+\cdots+r_d=k} \Pi_i a_i^{2r_i} C_{r_1, \cdots, r_d} \frac{2k!}{2r_1!\cdots 2r_d!} 
\end{equation*}
We now use the fact that the length of $a$ is $1$, as before, to obtain
\begin{equation*}
C_k \frac{k!}{r_1!\cdots r_d!} = C_{r_1, \cdots, r_d} \frac{2k!}{2r_1!\cdots 2r_d!} 
\end{equation*}
Finally, we use the fact that $x$ is also a unit vector to obtain
\begin{equation*}
\begin{split}
\sum_{r_1+\cdots +r_d=k} &C_{r_1, \cdots, r_d} \frac{k!}{r_1!\cdots r_d!}  = 1\\
 = &C_k\sum_{r_1+\cdots +r_d=k}  \frac{k!^2}{r_1!^2\cdots r_d!^2}\frac{2r_1!\cdots 2r_d!}{2k!}
\end{split}
\end{equation*}
Thus
\begin{equation}
C_k = \frac{1}{\sum_{r_1+\cdots +r_d=k} \frac{k!^2}{r_1!^2\cdots r_d!^2}\frac{2r_1!\cdots 2r_d!}{2k!}}
\end{equation}
And 
\begin{equation}
C_{s_1, \cdots , s_d} = \frac{1}{\sum_{r_1+\cdots +r_d=k} \frac{s_1!^2\cdots s_d!^2}{r_1!^2\cdots r_d!^2}\frac{2r_1!\cdots 2r_d!}{2s_1!\cdots 2s_d!}}
\end{equation}
Again, we provide a few specific values

\begin{table}[h!]

\begin{tabular}{|c|c|c|}
\hline
$k$  & Moment & Value \\
\hline 
$k=1$ & $\langle x_i^2\rangle$ & $\frac{1}{d}$\\
\hline
$k=2$ & $\langle x_i^4\rangle$ & $\frac{3}{d(d+2)}$\\
\hline
$k=2$ & $\langle x_i^2 x_j^2\rangle$ & $\frac{1}{d(d+2)}$\\
\hline
\end{tabular}

\end{table}

\section{Comparison of various measures of mixed state fidelity}
In this section, we will compare the various fidelity measures that can be used to quantify the error in mixed state preparation. Let us suppose that $\rho_{\text{targ}}$ is a $d-$dimensional ``target" quantum state which we intend to prepare.  Let $\rho_{\text{real}}$ be the realised state. There are several ways to quantify the magnitude of the error $\epsilon = \rho_{\text{targ}}-\rho_{\text{real}} $. 
\begin{itemize}
\item[i.] Frobenius norm/Hilbert-Schmidt norm/Schatten $2-$norm: $\sqrt{\text{Tr}(\epsilon^2)}=||\epsilon||_2$.
\item[ii.] Uhlmann fidelity: $\left(\text{Tr}\sqrt{\sqrt{\rho_{\text{targ}}} \rho_{\text{real}}\sqrt{ \rho_{\text{targ}}}}\right)^2$.
\end{itemize}

The three measures contain overlapping information. We evaluate the accuracy of $\rho_{\text{real}}$ based on observable values. If $\hat{O}$ is an observable, the error in its expectation value is $\text{Tr}(\epsilon \hat{O})$, which is a dot product and therefore, Cauchy-Schwarz inequality gives
\begin{equation*}
\text{Tr}(\epsilon \hat{O}) \leq ||\epsilon||_2 ||\hat{O}||_2
\end{equation*}
Thus in, case of a constant $\epsilon$, the Frobenius norm is straighforward to interpret.  However,  in real experiments,  $\epsilon$ is a random matrix and therefore, we need to construct averaged error measures. For instance, consider $\rho_{\text{targ}}=\frac{1}{d}\mathbbm{1}$ which can be approximately prepared using Haar random quantum states. That is,   $\rho_{\text{real}}=\frac{1}{\nu}\sum_{j=1}^{\nu} \ket{\psi_j}\bra{\psi_j}$ where $\{\ket{\psi_j}\}$ are Haar-random states and $\nu$ is the number of experimental shots.  In this case, $\epsilon = \frac{1}{d}\mathbbm{1} -\frac{1}{\nu}\sum_{j=1}^{\nu} \ket{\psi_j}\bra{\psi_j} $ depends on $\{\ket{\psi_j}\}$ and is therefore a random matrix, drawn from some underlying distribution.  The obvious candidate to quantify the error is $\langle ||\epsilon||_2^2\rangle $. We will however show that this is not the most appropriate measure. 

The fluctuation of the random matrix $\epsilon$ is characterized by the covariance matrix $\chi$ of $\epsilon$, which is the $d^2\times d^2$ matrix defined as
\begin{equation*}
\chi_{ij; kl} = \langle \epsilon_{ij}\epsilon_{lk} \rangle
\end{equation*}
Here, $\langle \cdot \rangle$ represents the average over the random $\epsilon$.   It follows that $\langle \epsilon^2\rangle$ is a partial trace of $\chi$. 
\begin{equation*}
\langle \epsilon^2\rangle _{ij} = \sum_k \chi_{ik;jk}
\end{equation*}
and
\begin{equation}\label{eps_to_chi}
\langle ||\epsilon||_2^2\rangle = \langle \text{Tr}(\epsilon^2)\rangle = \text{Tr}(\chi)
\end{equation}
The quantity of interest is the average error in measurements, given by the square average of  $\text{Tr}(\epsilon \hat{O})$
\begin{equation*}
\langle [\text{Tr}(\epsilon \hat{O})]^2\rangle =\sum \chi_{ij;kl}\hat{O}_{ij}\hat{O}_{kl} \leq \sigma_{\text{max}}(\chi)||\hat{O}||_2^2
\end{equation*}
Thus, the most relevant measure of error is given by the maximum singular value of $\chi$, $\sigma_{\text{max}}(\chi)$.  The average of the Frobenious norm, $\langle ||\epsilon||_2^2\rangle $, is in fact the sum of the singular values of $\chi$ (Eq.~\ref{eps_to_chi}) and is therefore much larger than the above limit. 
\begin{equation*}
\sigma_{\text{max}}(\chi) \leq \langle \text{Tr}(\epsilon^2)\rangle = \text{Tr}(\chi)
\end{equation*}
The equality holds for \textit{fixed} $\epsilon$. 

\subsection{Uhlmann Fidelity}

We now argue that Uhlmann fidelity is not the most appropriate measure for this purpose.  The Uhlmann fidelity is based on purification  or dilation of a mixed state into a pure state in a Higher dimensional space. For a given mixed state $\rho$ acting on $\mathbbm{C}^d$, one can find a pure state $\ket{\psi}\in \mathbbm{C}^{d^2}$ such that we recover $\rho$ after a partial trace: $\rho =\text{Tr}_1(\ket{\psi}\bra{\psi})$.  The fidelity between two states, $\rho_{\text{targ}}$ and $\rho_{\text{real}}$ is defined as
\begin{equation*}
F(\rho_{\text{targ}}, \rho_{\text{real}}) =\max |\langle \psi_{\text{targ}}| \psi_{\text{real}}\rangle|^2
\end{equation*}
Here, $\psi_{\text{targ}}$ and $\psi_{\text{real}}$ are purifications of $\rho_{\text{targ}}$ and $\rho_{\text{real}}$ respectively and the maximization is over all such purifications.  

The fidelity between two pure states represents the distinguishability between them. That is, it answers the question: how distinguishable are the states if one measures the same observable on them? For instance, for two states $\ket{\psi_1}$ and $\ket{\psi_2}$, the observable that best distinguishes them is $\hat{O}=\ket{\psi_1}\bra{\psi_1}-\ket{\psi_2}\bra{\psi_2}$. It's expectation value differs by $2(1-F)$ between the states, where $F=|\bra{\psi_1}\psi_2\rangle|^2$ is the fidelity. 

Therefore, the Uhlmann fidelity can be interpreted as a measure of the distinguishability of the \textit{purifications} of the mixed states. More precisely, it is a measure of the distinguishability of the least distinguishable purifications of the mixed state. While this may sound quite relevant for our purpose, this distinguishability is in the dilated space, which is not physically accessible. Often, the observable that distinguishes between the purifications is an entangling operator  on the \textit{dilated space} and therefore is not physically accessible. Our purpose here is to evaluate the distinguishability between the mixed states using observables acting on $\mathbbm{C}^d$.  Therefore, the Uhlmann fidelity, particlualrly while dealing with highly mixed states in Hiulbert spaces of large dimension, overestimates the error. 

We consider a simple example to illustrate the point. Let $\rho_{\text{targ}}=\frac{1}{d}\mathbbm{1}$ and $\rho_{\text{real}}=\frac{1}{\nu}\sum_{j=1}^{\nu}\ket{\psi_j}\bra{\psi_j}$ with $\bra{\psi_i}\psi_j\rangle =\delta_{ij}$. That is, we are approximating $\frac{1}{d}\mathbbm{1}$ by a mixture of $\nu$ orthonormal states with $\nu < d$. It is straightforward to show that 
\begin{equation}
\begin{split}
1-F(\rho_{\text{targ}}, \rho_{\text{real}})= \frac{d-\nu}{d}\\
||\rho_{\text{targ}}-\rho_{\text{real}}||^2_2 = \frac{d-\nu}{d\nu}
\end{split}
\end{equation}
Clearly, the Frobenius norm is much lower. In particular when $\nu=d/2$, $2(1-F)=1$ indicating that the purifications are quite distinguishable while $||\rho_{\text{targ}}-\rho_{\text{real}}||^2_2  = 1/d$, which approaches zero for large $d$.  However, we will show that the distinguishability represented by Uhlmann fidelity involves highly entangling observables in the dilated space and the Frobenious norm sets the relevant upper bound for observables within the original Hilbert space. 

Let $\{\ket{\psi_1}, \cdots, \ket{\psi_d}\}$ be a basis, completed from the states used to construct $\rho_{\text{real}}$. A choice of purification with minimal fidelity is
\begin{equation*}
\begin{split}
\ket{\psi_{\text{targ}}} =\frac{1}{\sqrt{d}}\sum_{j=1}^d \ket{\psi_j}\otimes \ket{\psi_j}\\
\ket{\psi_{\text{real}}} =\frac{1}{\sqrt{\nu}}\sum_{j=1}^{\nu} \ket{\psi_j}\otimes \ket{\psi_j}\\
\end{split}
\end{equation*}
The Observable that best distinguishes between them is $\hat{O}= \ket{\psi_{\text{targ}}} \bra{\psi_{\text{targ}}}-\ket{\psi_{\text{real}}}\bra{\psi_{\text{real}}} $ which is indeed highly entangling and cannot be measured within the original space.  In fact, the Frobenius norm in the dilated space is
\begin{equation*}
|| \ket{\psi_{\text{targ}}} \bra{\psi_{\text{targ}}}-\ket{\psi_{\text{real}}}\bra{\psi_{\text{real}}} ||_2^2 = 2(1-F)
\end{equation*}

\section{Convergence theorems}\label{convergence}

In this section, we prove theorem $1$ from the main text.  We begin with Haar random states to produce $\rho_{\text{targ}}=\frac{1}{d}\mathbbm{1}$.
\subsection{Haar random states}
If we pick $\nu$ states $\psi^{(r)}\in \mathbbm{C}^d$ with $r=1, \cdots, \nu$ and construct a mixed state, we obtain
\begin{equation*}
\rho_{\text{real}} = \frac{1}{\nu}\sum_r \ket{\psi^{(r)}}\bra{\psi^{(r)}}
\end{equation*}
It follows that $\langle \epsilon \rangle =0$. The covariance matrix is 
\begin{equation}
\chi_{ij;kl} = \frac{1}{\nu^2}\sum_{r, s} \langle z^{(r)*}_i z^{(r)}_j z^{(s)*}_l z^{(s)}_k\rangle  - \frac{1}{d^2}\delta_{ij}\delta_{kl}
\end{equation}
We use the formulae from section \ref{moments} to evaluate the above.
\begin{equation}
\begin{split}
\chi_{ij;kl} =& \frac{1}{\nu^2}\sum_{r} \langle z^{(r)*}_i z^{(r)}_j z^{(r)*}_l z^{(r)}_k\rangle\\
 +&\frac{1}{\nu^2}\sum_{r\neq s} \langle z^{(r)*}_i z^{(r)}_j \rangle \langle z^{(s)*}_l z^{(s)}_k\rangle  - \frac{1}{d^2}\delta_{ij}\delta_{kl}\\
 =& \frac{1}{\nu} \left(  \frac{\delta_{ij}\delta_{kl}+\delta_{ik}\delta_{jl}}{d(d+1)}\right) + \frac{\nu-1}{\nu}\frac{\delta_{ij}\delta_{kl}}{d^2}\\
 -& \frac{1}{d^2}\delta_{ij}\delta_{kl}
\end{split}
\end{equation}
Upon simplifying,
\begin{equation}
\chi_{ij;kl}  = \frac{\delta_{ik}\delta_{jl}}{\nu d(d+1)}-\frac{\delta_{ij}\delta_{kl}}{\nu d^2(d+1)}
\end{equation}
Thus,
\begin{equation}
\sigma_{\text{max}}(\chi) = \frac{1}{\nu d (d+1)}
\end{equation}

\subsection{Uncorrelated $\ell-$ bit states: Proof of convergence theorem 1}
We will now consider a system with a Hilbert space $(\mathbbm{C}^d)^{\otimes N}=\mathbbm{C}^{Nd}$. That is, $N$ copies of a $d-$dimensional quantum system. In the previous section, we studied the rate of convergence of a mixture of Haar random states to the uniformly mixed state. Here, we consider the rate of convergence of a mixture random product states to the uniform mixed state. That is, states  of the form $\ket{\psi}=\ket{\psi_1}\otimes \cdots \otimes \ket{\psi_{N}}$ where each $\ket{\psi_i}\in\mathbbm{C}^{d}$ is Haar-random.  We will compute the maximum singular value of $\chi$ for this case.

It is convenient to identify the singular vectors of $\chi$ for the $d-$dimensional subsystems.  $\chi$ acts on the space of $d\times d$ Hermitian matrices, which is a $d^2$ dimensional real vector space. Let us consider the operator $\frac{1}{\sqrt{d}}\mathbbm{1}$. This is a normalized vector in this space. We identify $d\times d$ Hermitian operators $X^{(1)}, \cdots, X^{(d^2-1)}$ which, together with $\frac{1}{\sqrt{d}}$ form an orthonormal basis.  That is, $\text{Tr}(X^{(\mu)})=0$ and $\text{Tr}(X^{(\mu)}X^{(\sigma)})=\delta_{\mu \sigma}$.  We show that these are the singular vectors of $\chi$, introduced in the previous section. To see this,
\begin{equation}
\begin{split}
\chi_{ij;kl}X^{(\mu)}_{kl} &= \frac{\delta_{ik}\delta_{jl}}{\nu d(d+1)}X^{(\mu)}_{kl}-\frac{\delta_{ij}\delta_{kl}}{\nu d^2(d+1)}X^{(\mu)}_{kl}\\
&=\frac{X^{(\mu)}_{ij}}{\nu d(d+1)}
\end{split}
\end{equation}
Moreover, $\chi \mathbbm{1} = 0$.  The state $\rho_{\text{real}}$ in this case is separable. That is, it is an average of product states of the $N$ subsystems.  Moreover, the averaging is over an uncorrelated distribution over the $N$ subsystems.  Therefore, we can compute the singular values of $\chi$ for $N$ systems using the singular values for the $\chi$ for each system. We invoke the moment matrix of $\rho_{\text{real}}$ for this purpose.  It is defined as
\begin{equation}
M_{ij; kl} = \langle \rho_{\text{real}, ij} \rho_{\text{real}, lk}\rangle
\end{equation}
The covariance matrix of $\rho_{\text{real}, ij}$ is the moment matrix of $\epsilon$, i.e., $\chi$ is related to $M$ as
\begin{equation}
\chi_{ij; kl} = M_{ij; kl}-\frac{\delta_{ij}\delta_{kl}}{d^2}
\end{equation}
The singular values of $M$ are $1/\nu d$ with singular vector $\frac{1}{\sqrt{d}}\mathbbm{1}$ and $\frac{1}{\nu d (d+1)}$ with singular vectors $X^{(\mu)}$ defined above. 
The moment matrix has an advantage over the covariance matrix because, in the case of uncorrelated distribution of the $N$ quantum systems,  the moment matrix of the full system is given by the tensor product of the moment matrices of the subsystems. That is, $M^{\otimes N}$. In other words, the covariance matrix  $\chi$ for the full system is
\begin{equation}
\chi = M^{\otimes N} - \frac{\mathbbm{1}}{d^{2N}}
\end{equation}
The singular values of the covariance matrix therefore are
\begin{equation}
\frac{1}{\nu d^k}\frac{1}{ (d (d+1))^{N-k}} =\frac{1}{\nu d^N (d+1)^{N-k}} \text{ for } k = 0, 1, 2, \cdots N-1
\end{equation}
The largest is $1/(\nu d^N(d+1))$. We can now use this to obtain the lagest singular value for $\ell-$bit states:
\begin{equation}
\frac{1}{\nu d^N (d^{\ell}+1)}
\end{equation}
\subsection{Correlated $\ell-$ bit states: Proof of convergence theorem 2}
We will now prove convergence theorem 2 from the main text for qudits.  If $\rho$ is a pure state of $\ell$ qudits, then the trace-free part $X=\frac{1}{d^\ell}\mathbbm{1}-\rho$ satisfies 
$$
\text{Tr}X^2 =1-\frac{1}{d^\ell}
$$
The residue $X$ is similar to the pauli-terms in the qubit case.  Thus, if $M_{ij;kl} = \rho_{ij}\rho_{lk}$, then $M$ has one singular value equal to $\frac{1}{d^\ell}$ and the remaining $d^{2\ell}-1$ of them sum up to $1-\frac{1}{d^\ell}$. That is, the maximum singular value after $\frac{1}{d^\ell}$ is at least $\frac{1}{d^{\ell}(d^\ell +1)}$.   Let $\rho^{(1)}\otimes \cdots \otimes \rho^{(r)}$ be a product state where each $\rho^{(j)}$ is an $\ell-$qudit state and $\epsilon= \frac{1}{d^N}\mathbbm{1}-\rho^{(1)}\otimes \cdots \otimes \rho^{(r)}$.  The corresponding $\chi$ matrix is $\langle M^{\otimes r} \rangle - \frac{1}{d^{2N}}\mathbbm{1}$. Regardless of the distribution (correlated or uncorrelated) of $\rho^{(j)}$, the largest singular value is at least $ \frac{1}{ d^N (d^\ell+1)}$ and thus, after $\nu$ samples,
\begin{equation}
\sigma_{\text{max}}(\chi)\geq \frac{1}{ \nu d^N (d^\ell+1)}
\end{equation}

\section{Sampling errors in tomography}
In this section, we will discuss how to estimate the total sampling error in process tomography experiments  involving a mixed initial state. In particular, we will derive Eq.~($1$).  We begin with a quick look at the sampling error in estimating classical probabilities.  Let $\{p_1, \cdots, p_k\}$ be classical probabilities associated with $k$ outcomes which we intend to estimate using $\nu$ samples. Of the $\nu$ samples, let us assume that $\nu_i$ of them correspond to the outcome $i$.  $\nu_1+\cdots + \nu_k = \nu$ and the maximum likelihood estimate of $p_i$ is $p_i'= \nu_i/\nu$. The sampling error in this estimation is given by
\begin{equation*}
\sum_i \left( \frac{\nu_i}{\nu}-p_i\right)^2 = \sum_i \frac{\nu_i^2}{\nu^2} +p_i^2 -2\frac{\nu_i p_i}{\nu}
\end{equation*}
We average this error over all possible outcomes $\{\nu_1, \cdots, \nu_k\}$, assuming that they are truly sampled from $\{p_1, \cdots, p_k\}$. The resulting error is
\begin{equation*}
\sum_i \frac{p_i(1-p_i)}{\nu} = \frac{1-\sum_i p_i^2}{\nu}
\end{equation*}
Thus,  the sampling error scales as $\sim 1/\sqrt{\nu}$.  However, in a quantum mechanical experiment, one can use correlated samples to decrease the sampling error. 

Let us now consider a typical measurement used in a process tomography. Let us assume that the initial state is mixed, $\rho_{\text{targ}}$ and the final state is $\rho=\Phi(\rho_{\text{target}})$. We intend to measure the final state in the basis $\{\ket{1}, \cdots, \ket{k}\}$. The probabilities of the $k$ outcomes are $p_i = \bra{i}\Phi(\rho_{\text{target}})\ket{i}$, which we estimate as $p_i'$ using $\nu$ samples. If we use uncorrelated samples, we can estimate $p_i$ within an error of $1/\sqrt{\nu}$ \textit{assuming} that $\rho_{\text{targ}}$ has been prepared perfectly. However, practically a mixed state is prepared by generating controlled samples of pure states that average down to the target mixed state, there will be a non-zero sampling error in the state.  For instance, if the real state is $\rho_{\text{real}}$, the measured values $p'_i$ actually estimate $p''_i = \bra{i}\Phi(\rho_{\text{real}})\ket{i}$ with a sampling error.  Thus, the total error is
\begin{equation*}
\sum(p_i'-p_i)^2  = \sum (p_i - p_i'' + p_i'' - p_i')^2
\end{equation*}

Note that $(p_i-p''_i)$ and $(p_i''-p_i')$ are \textit{independent } random variables: one corresponds to the sampling error at the state preparation and the other corresponds to the sampling error at measurement. Upon averaging, the total error is
\begin{equation*}
\langle \sum(p_i'-p_i)^2 \rangle  = \langle \sum(p_i-p_i'')^2 \rangle + \langle \sum(p_i''-p_i')^2 \rangle
\end{equation*}
The first term is the sampling error at state preparation, related to $\Phi(\rho_{\text{real}}-\rho_{\text{target}})$ and the second is the sampling error at measurement.  From theorem $1$, it follows that the first term also scales as $\sim 1/\sqrt{\nu}$ if we use uncorrelated initial states.




\end{document}